\RequirePackage{fix-cm}
\documentclass{svjour3}                     
\smartqed  

\usepackage[table,xcdraw]{xcolor}
\usepackage{hyperref}
\usepackage{caption}
\usepackage{subcaption}
\usepackage{graphicx}
\usepackage{tcolorbox}
\usepackage{tabularx}
\usepackage{lipsum}
\usepackage{balance}
\usepackage{blindtext,graphicx}
\usepackage[absolute]{textpos}
\usepackage[rightmargin=0.5pt,leftmargin=0.5pt]{quoting}
  
  \newenvironment{myquote}%
  {}

\usepackage{multirow}
\usepackage{longtable}
\usepackage{tabu}
\usepackage{makecell}
\usepackage{fontawesome}
\usepackage{csquotes}
\usepackage{relsize,etoolbox}
\usepackage{booktabs}
\usepackage{lscape}
\usepackage{etoolbox}
\newcommand*{\affaddr}[1]{#1} 
\newcommand*{\affmark}[1][*]{\textsuperscript{#1}}

\newcommand{\todo}[1]{}
\renewcommand{\todo}[1]{{\color{red} TODO: {#1}}}
\begin{document}
\sloppy

\title{Automated Detection, Categorisation and Developers' Experience with the Violations of Honesty in  Mobile Apps}



\author{Humphrey O. Obie \protect\affmark[1]         \and
        Hung Du \affmark[2] 
        \and Kashumi Madampe \affmark[1]
        \and Mojtaba Shahin \affmark[3]
        \and Idowu Ilekura \affmark[4]
        \and John Grundy \affmark[1]
        \and Li Li \affmark[5]
        \and Jon Whittle \affmark[6]
        \and Burak Turhan \affmark[7]
        \and Hourieh Khalajzadeh \affmark[2]
}


\institute{ \and
              Humphrey O. Obie \\
              \email{humphrey.obie@monash.edu}             \\
           \and
           Hung Du\\
              \email{hung.du@deakin.edu.au}             \\
           \and
           Kashumi Madampe \\
              \email{kashumi.madampe@monash.edu}             \\
           \and
            Mojtaba Shahin\\
              \email{mojtaba.shahin@rmit.edu.au} \\
              \and
            Idowu Ilekura \\
              \email{ilekuraidowu@gmail.com}             \\
              \and
            John Grundy\\
              \email{john.grundy@monash.edu} \\
           \and
           Li Li\\
              \email{lilicoding@ieee.org} \\
              \and
           Jon Whittle\\
              \email{jon.whittle@data61.csiro.au} \\
              \and
            Burak Turhan\\
              \email{burak.turhan@oulu.fi} \\
              \and
           Hourieh Khalajzadeh \at
             \email{hkhalajzadeh@deakin.edu.au} \\
              \affaddr{\affmark[1]Monash University, Melbourne, Australia}\\
\affaddr{\affmark[2]Deakin University, Melbourne, Australia}\\
\affaddr{\affmark[3]RMIT University, Melbourne, Australia}\\
\affaddr{\affmark[4]Data Science Nigeria, Lagos, Nigeria}\\
\affaddr{\affmark[5]School of Software, Beihang University, China}\\
\affaddr{\affmark[6]CSIRO’s Data61, Melbourne, Australia}\\
\affaddr{\affmark[7]University of Oulu, Oulu, Finland}
}

\date{Received: date / Accepted: date}

\maketitle

\begin{abstract}
Human values such as honesty, social responsibility, fairness, privacy, and the like are things considered important by individuals and society. Software systems, including mobile software applications (apps), may ignore or violate such values, leading to negative effects in various ways for individuals and society. While some works have investigated different aspects of human values in software engineering, this mixed-methods study focuses on \emph{honesty} as a critical human value. 
In particular, we studied (i) how to detect honesty violations in mobile apps, 
(ii) the types of honesty violations in mobile apps, and (iii) the perspectives of app developers on these detected honesty violations. We first develop and evaluate 7 machine learning (ML) models to automatically detect violations of the value of honesty in app reviews from an end-user perspective. The most promising was a Deep Neural Network model with F1 score of 0.921.  We then conducted a manual analysis of 401 reviews containing honesty violations and characterised honest violations in mobile apps into 10 categories: unfair cancellation and refund policies; false advertisements; delusive subscriptions; cheating systems; inaccurate information; unfair fees; no service; deletion of reviews; impersonation; and fraudulent-looking apps. A developer survey and interview study with mobile developers then identified 7 key causes behind honesty violations in mobile apps and 8 strategies to avoid or fix such violations. The findings of our developer study also articulate the negative consequences that honesty violations might bring for businesses, developers, and users. Finally, the app developers’ feedback shows that our prototype ML-based models can have promising benefits in practice. 
\end{abstract}

\section{Introduction}
\label{sec:introduction}
Human values, such as \textit{integrity, privacy, curiosity, security,} and \textit{honesty}, are the guiding principles for what people consider important in life \cite{Cheng:2010}. These values influence the choices, decisions, relationships, and the concept of ethics for people and society at large whether or not they are formally articulated in this terminology \cite{Schwartz:1992}. The relationship between human values and technologies is important, especially for ubiquitous technologies like mobile software applications (apps) \cite{Obie:2020}. Mobile apps are a convenience to modern society and have seen usage in carrying out both simple and complex tasks, from entertainment (e.g., video sharing apps) and health (e.g., fitness trackers) to finance (e.g., banking apps).  End-users of these apps hold certain expectations influenced by their human values considerations, e.g., the privacy of data, transparency of processes in apps, and ethical behaviour of platforms and software companies \cite{Obie:2020}. The violation of these value considerations is detrimental to the end-user, software platforms, companies, and society in general \cite{Whittle:2021}.

Recent work on human values in software engineering (SE), based on the Schwartz theory of basic human values \cite{Schwartz:1992,Schwartz:2012}, has mapped human values to specific ethical principles. For example, Perera et al. mapped values to the GDPR principles \cite{Perera:2019} and Winter et al. mapped values to the ACM Code of Ethics \cite{Winter:2018}. Other studies such as \cite{Obie:2020} have explored the violation of human values in mobile apps using app reviews as a proxy. The recent study by Obie et al. showed that the value of \textit{honesty} (a sub-item of \textit{benevolence} based on Schwartz theory \cite{Schwartz:1992}) is violated by mobile apps \cite{Obie:2020}.

Honesty, often perceived to be a very important human value \cite{MillerChristianB2021HTPa}, describes a character quality of being sincere, truthful, fair, and straightforward, and refraining from lying, cheating, deceit, and fraud \cite{dictionary:2021}. The importance of the value of honesty is clearly articulated in the ACM Code of Ethics: \emph{``Honesty is an essential component of trust. A computing professional should be transparent and provide full disclosure of all pertinent system limitations and potential problems. Making deliberately false or misleading claims, fabricating or falsifying data, and other dishonest conduct are a violation of the Code.''} \cite{acmcode:2021}. Nonetheless, there have been many flagrant violations of the value of honesty by mobile app platforms and software companies in ways that are collectively called \textit{dark patterns} \cite{dong2018frauddroid,van2019development,hu2019dating,samhi2022difuzer,gao2022demystifying}. These dark patterns, a violation of the value of honesty, entail sophisticated design practices that can trick or manipulate consumers into buying products or services or giving up their privacy \cite{Gizmodo:2022}. Furthermore, some of these dark patterns and honesty violations have been flagged in a recent report by the Federal Trade Commission (FTC) \cite{FTC:2022}. Other examples include companies deliberately hiding data breaches from the authorities and customers \cite{Bowman:2021,Shaffery:2021}. These violations of the value of honesty result in decreased trust from users, poor uptake of apps, and reputational and financial damage to the organisations involved. This also emphasises the need to consider human values more proactively in software engineering practice. 

\textcolor{black}{This work aims to gain a comprehensive understanding of honesty violations in mobile apps. To this end, we conducted a mixed-method study. Given that user's comments expressed in app reviews have been shown to be a proxy for detecting users' challenges and requirements \cite{AlOmar:2021,Obie:2020,shams2020society,DiSorbo:2016,Guzman:2014}, we first developed and evaluated seven machine learning models to learn the features that are representative of the violation of honesty in app reviews. The best-performing model (a Deep Neural Network) has an F1 score of $0.921$, a precision of $0.911$, and a recall of $0.932$. Beyond the automatic detection of honesty violations, we then manually analysed 401 reviews containing honesty violations. Our resulting taxonomy shows that honesty violations can be characterised into ten categories: \textbf{unfair cancellation and refund policies, false advertisements, delusive subscriptions, cheating systems, inaccurate information, unfair fees, no service, deletion of reviews, impersonation,} and \textbf{fraudulent-looking apps}. Finally, we conducted a developer study\footnote{Approved by Monash Human Research Ethics Committee. Approval Number: 35070} with mobile app developers to explore their experiences with honesty violations. The analysis of qualitative data from the developer study resulted in identifying a set of causes (business, developer, app platform, user and competitor drivers), parties responsible for the honesty violations (product owners, managers, business analysts, user support roles -- in addition to developers, and business), consequences of honesty violations on businesses, app developers, and end users (e.g., bad reputation for the company, developers experiencing negative emotions, and identity theft of users), and strategies the app developers use to avoid (e.g., strengthening designing practices) and fix (e.g., thoroughly investigating the violation and fixing it) honesty violations in the apps they develop. The developer study also shows that the automatic detection of honesty violations in app reviews bring several benefits to business, developers, app platforms, and users. From this point onwards, the terms ``mobile app developer'' and ``developer'' are used interchangeably.}

We published preliminary results of our machine learning-based classification work at the Mining Software Repositories (MSR) conference in 2022 \cite{MSR2022Humphrey}. We substantially extend this previously published work here by (i) the inclusion of two new machine learning models, deep neural network (DNN) and generative adversarial network (GAN), with the new DNN replacing the prior support vector machine (SVM) as the best model; (ii) evaluation of these new models; and (iii) adding a new research question (\textbf{RQ3}). \textbf{RQ3} includes four sub-questions to seek mobile developers' views on the causes behind honesty violations in mobile apps (\textbf{RQ 3.1}), the potential consequences of honesty violations (\textbf{RQ3.2}), possible solutions to avoid or fix such violations (\textbf{RQ3.3}), and potential benefits of automated identification of honesty violations in mobile apps (\textbf{RQ3.4}). 



This work makes the following key contributions:

\begin{itemize}
    \item We present several machine learning models and datasets to aid the automatic detection of the violation of the human value of honesty in app reviews. Our publicly available replication package supports researchers and practitioners to adapt, replicate, and validate our study \cite{onlinedataset}.

\item We provide insights into the different categories of honesty violations prevalent in app reviews by creating a taxonomy based on a manual analysis of the honesty violations dataset.

\item We survey 70 app developer practitioners and interview 3 practitioners to get their feedback on the prevalence of honesty violations in their mobile apps, the causes of these issues, and feedback on our proposed machine learning-based classifier to help identify such violations from user app reviews.

\item We present an actionable framework for developers which gives a better understanding of the causes and consequences of honesty violations and strategies that can be used to avoid and fix honesty violations. 

\item We present a set of practical implications and future research directions to deal with the challenges of the violations of the human value of honesty in apps that would benefit end-users and society.

\end{itemize}

The rest of the paper is organised as follows: Section \ref{sec:relatedwork} summarises the related studies. In Section \ref{sec:reseachdesign}, we elaborate on the research design.  The findings of this study are reported in Sections \ref{sec:automaticclassification}, \ref{sec:categorieshonesty}, and \ref{sec:RQ3} for different research questions. Section \ref{sec:discussion} reflects on the findings and provides implications, followed by reporting on possible threats in Section \ref{sec:ThreatsValidity}. We conclude the paper in Section \ref{sec:Conclusion}.






 

   


\section{Motivating Examples}

Consider an example review of dubious charges to a user account for a calendar reminder app:
\textit{``I've been charged \$45+ on 2 separate occasions in the month I've had the `premium' version. It advertises \$3.50 for a premium subscription but saw nowhere where it said they would make additional charges.} 
Such reviews are common for many subscription-based apps. They are also very common for apps with optional premium versions where users find themselves unwittingly signed up to the premium charges. Many end users perceive these as deliberate attempts by app providers to dishonestly make money.
Some companies have been convicted of such dishonest practices. For example, Shaw Academy offered users a free trial to its online education platform and charged them a subscription fee even if they had cancelled before the end of the trial period and refused to refund the users \cite{ACS:2021}. The outcome of an investigation by the Australian Competition \& Consumer Commission  (ACCC) ordered the company to refund approximately $\$50,000$ to the affected users and pledge to improve their system \cite{ACS:2021}. 

Consider another example of dishonesty from a dating app. The dating platform (\url{Match.com}) has been accused of faking love interests using bots and fake profiles to fool consumers into buying subscriptions and exposing them to the risk of fraud and other deceptive practices \cite{TechCrunch:2019}. During a period of over three years, the company allegedly delivered marketing emails (i.e., the \textit{“You have caught his eye”} notification) to potential consumers after the company's internal system had already flagged the message sender as a suspected bot or scammer. The company also violated the “Restore Online Shopper's Confidence Act” (ROSCA) by making the unsubscription process tedious. Internal company documents showed that users need to make more than six clicks to cancel their subscription. This resulted in the U.S. Federal Trade Commission (FTC) suing \url{Match.com} for “deceptive advertising, billing, and cancellation practices” \cite{TechCrunch:2019}.

\section{Related Work} \label{sec:relatedwork}

\textbf{Mining App Reviews: } Many works have provided insights into app user reviews and how these reviews can aid software professionals in app requirements, design, maintenance \cite{Pelloni:2018,Carreno:2013,Seyff:2010} and evolution \cite{Adelina:2017,Li:2018,Li:2010,Palomba:2015}. Guzman and Maalej adopted Natural Language Processing (NLP) techniques to locate fine-grained app features in reviews with the aim of supporting software  requirements tasks \cite{Guzman:2014}. A related work utilised Latent Dirichlet Allocation (LDA) technique and linguistic rules to group feature requests from users as expressed in their reviews, and the results from this study showed that users care about frequent updates, improved support, more customisation options, and new levels (for game apps) \cite{Iacob:2013}.

Some studies have focused on the automatic classification of app reviews into useful categories. To aid software professionals in prioritising accessibility issues, AlOmar et al. developed a machine learning model for identifying accessibility-related complaints in app reviews \cite{AlOmar:2021}. Panichella et al. introduced a taxonomy for classifying app reviews and, using a combination of NLP and sentiment analysis, classified app reviews into their proposed taxonomy \cite{Panichella:2015}. 

Other works have introduced tools to support the extraction of insights from app reviews. For example, Vu et al. proposed MARK, a keyword-based tool for detecting trends and changes that relate to occurrences of serious issues in reviews \cite{Phong:2015}. Similarly, Di Sorbo et al. introduced SURF,  a tool that condenses thousands of reviews into coherent summaries to support change requests and planning of software releases \cite{DiSorbo:2016}. Our own work has classified various app reviews into different human value violations \cite{Obie:2020,obie2021does}, human-centric issues discussed in app reviews \cite{mathews2021ah,khalajzadeh2022supporting}, and a myriad of problems users have with eHealth apps \cite{haggag2022large}.

The above studies show that app reviews are a useful resource for gathering requirements, detecting issues, and more generally supporting software professionals in evolving their apps. This work also aims to support app maintenance and evolution by effectively detecting potential violations of the value of honesty from the user's perspective in app reviews. In addition, it would aid software professionals in delivering software products that build trust in users, as the honesty (real or perceived) of companies can affect how users engage with their products \cite{Zhu:2021}.

\textbf{Human Values in Software Engineering (SE): }
Human values are enduring beliefs that a specific mode of conduct or end state of existence is personally or socially preferable to an opposite or converse mode of conduct or end state of existence \cite{Rokeach:1973}. Human values have been well-studied in the social sciences and have begun to see adoption in other fields, including design \cite{Aldewereld:2015} and software engineering (SE) \cite{Mougouei2020engineering,li2021step}.

The study of human values in SE is a relatively nascent line of research \cite{Perera:2020,Mougouei:2018} and is mostly based on the widely accepted and adopted Schwartz theory of basic human values \cite{Schwartz:1992,Schwartz:2012}. The Schwartz theory is built on a survey conducted in over 80 countries covering different demographics. This theory categorises values into 10 broad categories, namely: \textit{self-direction, stimulation, hedonism, achievement, power, security, conformity, tradition, benevolence,} and \textit{universalism}. These 10 categories, in turn, are made up of 58 value items, e.g., the value category of \textit{\textbf{benevolence}} covers the value items of \textit{\textbf{honesty}, responsible, helpful, forgiving, loyal, mature love, a spiritual life, meaning in life,} and \textit{true friendship} (c.f \cite{Schwartz:1992}). However, our focus in this work is on the value item of \textit{honesty}, based upon the prevalence of the value category of \textit{benevolence} in prior research \cite{Obie:2020}, the recent cases of the violations of \textit{honesty} by companies in the media, e.g., \cite{TechCrunch:2019,ACS:2021}, and the need to understand this phenomenon more closely in SE.

Studies in the social sciences have investigated the value of (dis)honesty at the individual and organisational levels \cite{FOCHMANN2021250}, and the policy implication of dishonesty in everyday life \cite{Mazar:2006}; while others have explored the motivation for dishonest behaviours \cite{prooijen_lange_2016} including students in classroom settings \cite{lang2013cheating} and workers in crowd-working environments \cite{JacquemetNicolas2021Dtoi}. Keyes argues that euphemising the violation of the value of honesty desensitises people to its implications and consequences in society \cite{KeyesRalph2004Tpe}.

However, within the context of SE, Whittle et al. argued that software companies need to consider human values in the development of software systems and make them “first-class” entities throughout the software development life cycle \cite{Whittle:2021}. Another study made a case for the evolution of current software practices and frameworks to embed human values in technology, instead of a revolution of the SE field \cite{Hussain:2020}.

Another line of research considered methods for measuring human values in SE. For example, Winter et al. introduced the Values Q-sort instrument for measuring human values in SE \cite{Winter:2018}. Applying the Values Q-sort instrument to 12 software engineers resulted in 3 software engineer values “prototype”. Similarly, Shams et al. utilised the portrait values questionnaire (PVQ) to elicit the values of 193 Bangladeshi female farmers in a mobile app development project \cite{Shams:2021}. The result of the study showed that conformity and security were the most important values, while power, hedonism, and stimulation were the least important. More recently, Obie et al. argued that the instruments for eliciting and measuring values should be contextualised to specific domains \cite{obie2021does}.

\textbf{App Reviews and Human Values:} Recent studies have adopted the use of app reviews as an auxiliary data source for eliciting values requirements. Shams et al. analysed 1,522 reviews from 29 agricultural mobile apps to understand the values that are both represented and missing from these apps \cite{shams2020society}. Obie et al. proposed a keyword dictionary-based NLP classifier to detect the value categories violated in app reviews \cite{Obie:2020}. The results of the application of the classifier to 22,119 reviews showed that benevolence and self-direction were the most violated categories, while conformity and tradition were the least violated. 

Related works such as \cite{shams2020society,Obie:2020} have provided insights to violations of value categories. Our work complements these by zooming in on a specific value item; \textbf{\textit{honesty}} (within the most violated category of \textbf{\textit{benevolence}} \cite{Obie:2020}), to provide a more nuanced understanding of its violations. In addition,
we provide a taxonomy of the different categories of honesty violations in reviews to better understand how the violation of the value of honesty is reported. Our practitioner survey and interviews suggest automated identification of honesty violations from app reviews would be practically useful. We hope that other researchers would be encouraged to investigate other specific value categories, their discussion of violations by users in app reviews, and more generally to explore the field of human values in SE.

\section{Research Design}\label{sec:reseachdesign}

\textcolor{black}{Our goal in this study is to develop a deep understanding of honesty violations in mobile apps by automatically identifying reviews discussing honesty values, categorising the types of honesty violations, and exploring the perspectives of app developers about their perspectives on such honesty violations. To do this, we have formulated the following research questions that we need to answer (RQs):}




\begin{itemize}

\item \textbf{RQ1.} \textit{Can we effectively identify reviews documenting honesty violations automatically?} 
\textcolor{black}{We formed a large labelled dataset of app reviews and then trained a variety of machine learning classifiers to answer this RQ. Our best-performing classifier has an F1 score of 0.91.}

\item \textbf{RQ2.} \textit{What types of honesty violations are reported in these app reviews?}
\textcolor{black}{We manually inspected a sample of 401 honesty violation reviews and classified the honesty violations represented by each into ten distinct categories.}

\item \textbf{RQ3.} \textcolor{black}{ \textit{What is app developers’ experience with honesty violations in the mobile apps they develop and their perspective on automatic detection of honesty violations?}}
\textcolor{black}{We developed three subquestions to answer this RQ. We use in-depth interviews and a broad survey with the participation of 73 mobile app practitioners.}

\subitem \textbf{RQ3.1.}\textcolor{black}{\textit{What are the causes of honesty violations in mobile apps, and who is responsible for them?}}
\textcolor{black}{We want to know, according to developers’ experience with honesty violations in mobile apps they develop, what causes these honesty violations in mobile apps and who is responsible for them.}
 
\subitem \textbf{RQ3.2.}\textcolor{black}{\textit{What are the consequences of honesty violations in mobile apps on the end users and app developers/owners according to developers’ experience?}}
\textcolor{black}{The goal of this RQ is to understand the impacts of honesty violations on end users, and the developers themselves/owners of the mobile apps, as experienced by the mobile app developers.}

\subitem \textbf{RQ3.3.} \textcolor{black}{\textit{What strategies do developers use to handle honesty violations in mobile apps?}}
\textcolor{black}{This RQ aims to identify what strategies the mobile app developers use to avoid and/or fix reported honesty violations in mobile apps (or if they indeed do so).}

\subitem \textbf{RQ3.4.} \textit{What are the benefits of automatically detecting honesty violations in mobile apps?} 
Through this research question, we target exploring the potential benefits of automatic detection of honesty violations.

\end{itemize}

\begin{figure}
    \centering
    \includegraphics[width=\textwidth]{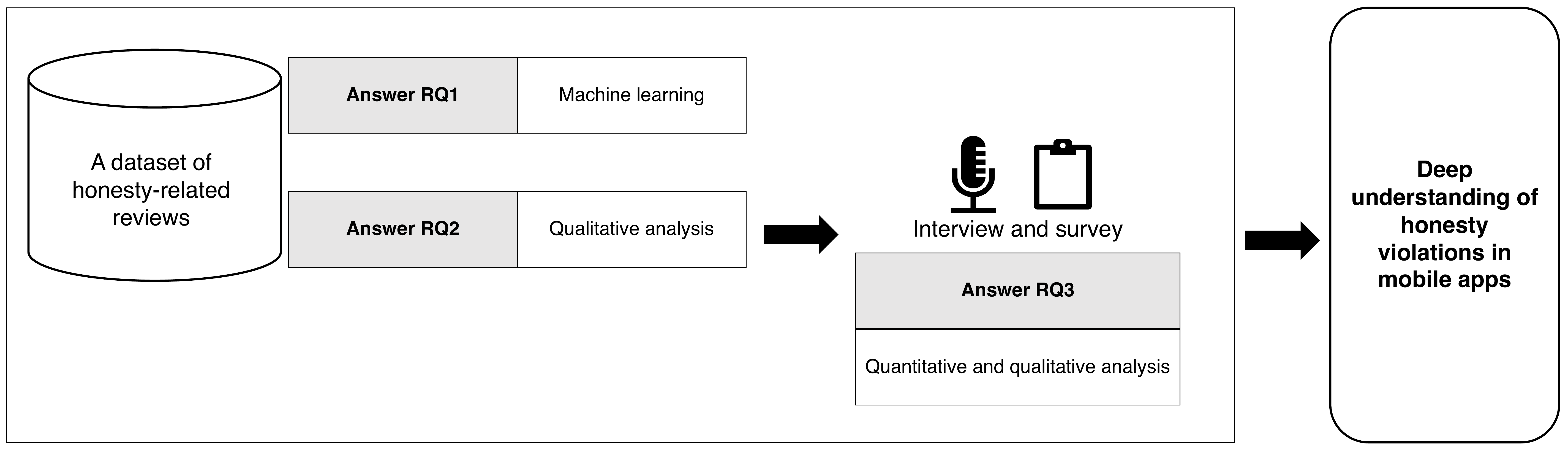}
    \caption{High-Level Overview of Mixed-methods Approach}
    \label{fig:high_level}
\end{figure}

\textbf{Approach for answering the above research questions. }A high-level overview of our mixed-methods approach is given in Fig. \ref{fig:high_level} and elaborated in figures and text in the respective sections answering the research questions.

\section {Automatic Classification of Honesty Violations (RQ1)} \label{sec:automaticclassification}

\begin{figure*}[t]
    \centering
    \includegraphics[width=0.95\textwidth]{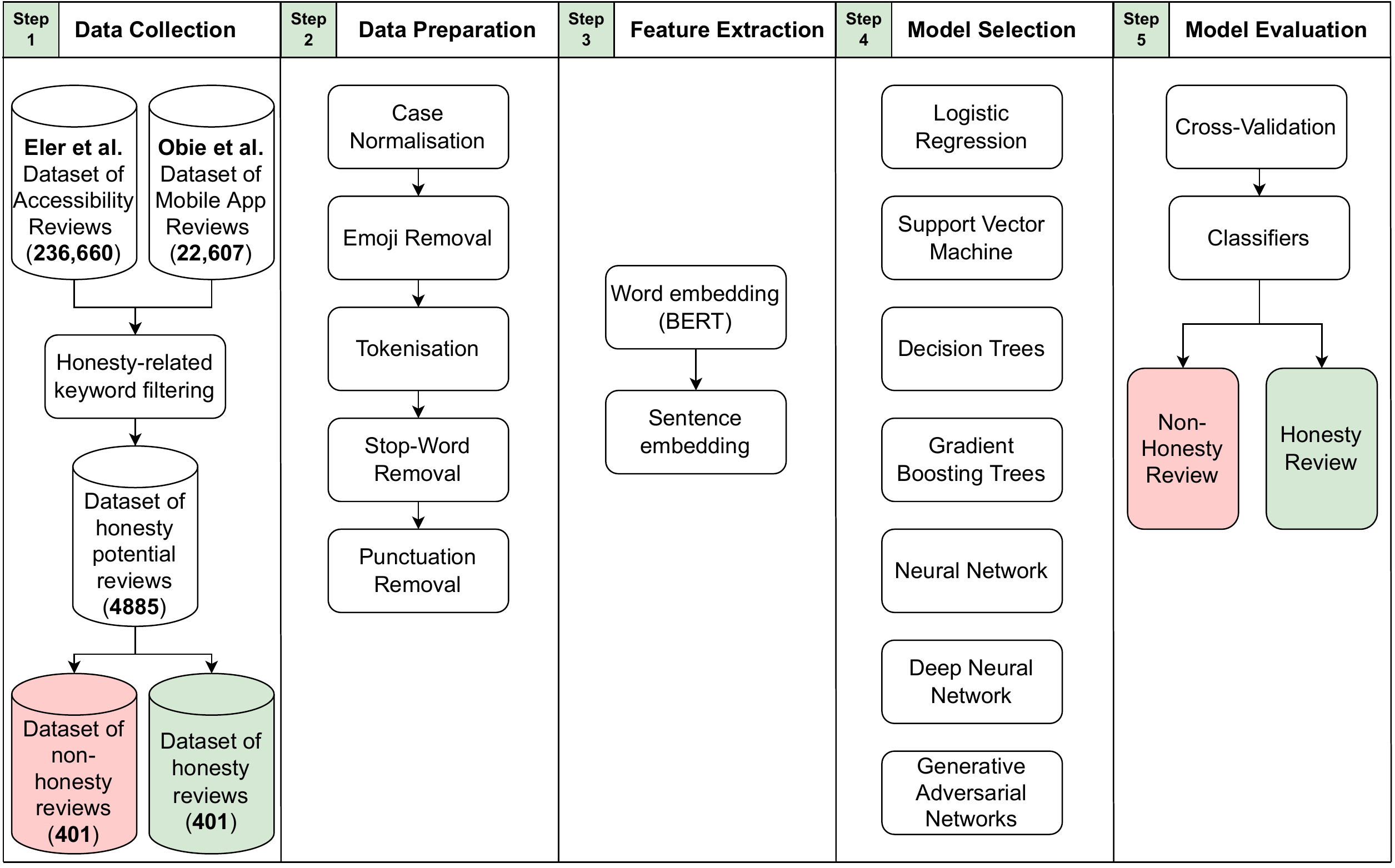}
    \caption{Honesty app review classification process} \label{fig:ml_process}
\end{figure*}

\subsection{A Dataset of Honesty-related Reviews}
\label{sec:dataset}
Our first step for answering RQ1 and RQ2 is creating a dataset of user reviews documenting perceived honesty violations by apps.
\subsubsection{Data Collection} To build this dataset, we collected a total of 236,660 reviews - 214,053 reviews from the public dataset of Eler et al. \cite{Eler:2019}, and an additional 22,607 reviews from the public dataset of Obie et al. \cite{Obie:2020}. These reviews were collected from a total of 713 apps in 25 categories. The apps and reviews were intended to cover a diverse range of categories and audiences. Table \ref{tab:data_statistics} summarises the statistics of our combined app review dataset. Our dataset can be found here \cite{onlinedataset}.

\begin{table}
\centering
\small
\caption{Statistics of the dataset.}
\label{table:API frequency for categories}
\begin{tabular}{|l|l|} 
\hline
Number of Apps  & 713      \\ 
\hline
App Categories  & 25       \\ 
\hline
All Reviews     & 236,660  \\ 
\hline
Honesty-related Reviews (after keywords filter) &   4,885\\
\hline
Honesty Violation Reviews (after manual validation) & 401 \\
\hline
\end{tabular}
\label{tab:data_statistics}
\end{table}

\subsubsection{Data Labeling} Given the sheer size of the dataset and the manual labour required to label the dataset, we used two approaches to label the 236,660 reviews: a keyword-based approach and manual labelling. We first adopt a set of keywords to filter the 236,660 reviews to include those related only to the value of honesty. These keywords are based on the dictionary of human values created by Obie et al. \cite{Obie:2020}. The set of keywords comprises a total of 48 words semantically related to honesty. The keywords are available in \cite{onlinedataset}. 
After applying this keyword filter, the number of reviews was reduced from 236,660 reviews to 4,885 potential candidate honesty-related reviews (we call these 4,885 reviews \textit{\textbf{honesty}\textbf{\_}\textbf{potential}} reviews). 

However, adopting a keyword-based approach is error-prone and may result in a lot of false positives. Hence, we manually analysed the \textit{\textbf{honesty}\textbf{\_}\textbf{potential}} reviews to exclude the false positives. The application of keywords filter and subsequent manual analysis check have been applied in recent studies \cite{Eler:2019,AlOmar:2021}.

The \textit{\textbf{honesty}\textbf{\_}\textbf{potential}} reviews were labelled and validated in $25\%$ increments in the following manner. The first analyst labelled the first $25\%$ percent of the \textit{\textbf{honesty}\textbf{\_}\textbf{potential}} reviews to determine which of the reviews contain the violation of the value of honesty as perceived by the user in the review. The second analyst validated the outcome. The disagreements were resolved in a meeting using the negotiated agreement approach to address issues of reliability \cite{2013_campbell_coding,1974_morrissey_sources}. Then the next $25\%$ were labelled by the first analyst, validated by the second analyst, and disagreements resolved in a meeting as in the first round. The same procedure was repeated for the third and fourth rounds of the labelling process. Also, the labelling and validation were done over eight weeks to avoid fatigue. Based on our manual labelling, we found that out of the $4,885$ filtered reviews (the \textit{\textbf{honesty}\textbf{\_}\textbf{potential}} reviews), only $401$ were honesty violations reviews, i.e., true positives. We refer to these $401$ honesty violations reviews as \textit{\textbf{honesty}\textbf{\_}\textbf{violations}} reviews.

Next, we randomly selected $401$ reviews from the remaining 4,484 \textit{\textbf{honesty}\textbf{\_}\textbf{potential}} reviews (4,885 \textit{\textbf{honesty}\textbf{\_}\textbf{potential}} reviews - 401 \textit{\textbf{honesty}\textbf{\_}\textbf{violations}} reviews). We refer to these 401 reviews, which contain honesty-related keywords (but not violations), as \textit{\textbf{honesty}\textbf{\_}\textbf{non}\textbf{\_}\textbf{violations}} reviews. We used a total of 802 reviews: 401 \textit{\textbf{honesty}\textbf{\_}\textbf{violations}} and 401 \textit{\textbf{honesty}\textbf{\_}\textbf{non}\textbf{\_}\textbf{violations}} reviews to build a balanced dataset called \textbf{honesty}\textbf{\_}\textbf{discussion} dataset for training and evaluating machine learning models in Section \ref{sec:automaticclassification}. We note here that using the manually validated false-positive \textit{honesty\_non\_violations} reviews is important for machine learning models. It is because these reviews include certain keywords syntactically related to honesty but semantically irrelevant to honesty violations - an important difference we want our models to learn. In summary, the \textbf{honesty}\textbf{\_}\textbf{discussion} dataset consists of 802 reviews: 401 \textit{\textbf{honesty}\textbf{\_}\textbf{violations}} reviews and 401 \textit{\textbf{honesty}\textbf{\_}\textbf{non}\textbf{\_}\textbf{violations}} reviews. Other studies have used similar numbers of text documents in classification tasks \cite{Levin:2017,Levin:2019}.


\subsection{Classification Approach} Manually classifying honesty violations in app reviews is challenging for practitioners because it is error-prone, labour-intensive, and demands substantial domain expertise. Hence, an automated approach is required to recognise honesty violations in app reviews. This research question aims to develop machine learning models to differentiate between honesty and non-honesty reviews automatically. As shown in Figure \ref{fig:ml_process}, the machine learning models are applied on the 802 \textbf{honesty}\textbf{\_}\textbf{discussion} dataset which consists of 401 \textit{\textbf{honesty}\textbf{\_}\textbf{violations}} reviews and 401 \textit{\textbf{honesty}\textbf{\_}\textbf{non}\textbf{\_}\textbf{violations}} reviews.

\subsubsection{Data Preparation} \label{sec:datapreparation}
We applied some common techniques to remove possible noise from the \textbf{honesty}\textbf{\_}\textbf{discussion} dataset. This step was needed so a learning model can classify reviews correctly. To achieve this, we applied natural language processing techniques such as removing capitalisation, removing emojis, tokenising, removing stop words, and removing punctuation.

\textbf{Case Normalisation:} is the process of transforming original review texts into their lowercase. This type of text cleansing helps us avoid repeated features of the same words with different font cases (e.g., ``Honesty" and ``honesty"). Furthermore, converting the text into its lowercase does not affect its context as well as the users' expressions in our scenario. 

\textbf{Emoji Removal:} Emojis are icons or a few Unicode characters that allow users to convey ideas, concepts, and emotions. 
If emojis are not carefully preprocessed, they can potentially affect the performance of a model in terms of accuracy. Hence, we removed emojis from the review texts.

\textbf{Tokenisation:} is the process of splitting each original text into a set of words that do not contain white space. We divided apps reviews into their constituent set of words.

\textbf{Stop-Word Removal:} Stop words such as \textit{is}, \textit{am}, \textit{are}, \textit{for}, \textit{the}, and others do not contain the conceptual meaning of a review and create noise for a classification model. Removing stop words from the review texts helps us avoid repeated features of the same phases (e.g., ``the bank account" and ``bank account"). In our experiment, we used a comprehensive set of stop words that are well-known to the natural language processing community\footnote{The stop words can be accessed at \url{https://gist.github.com/sebleier/554280\#gistcomment-3126707}}.

\textbf{Punctuation Removal:} We observed many reviews in the data collection containing punctuation such as \textit{"..., ??, :(,"} and others that do not significantly contribute to a classification model. Hence, we removed punctuation from the app reviews.

\subsubsection{Feature Extraction}\label{sec:featureexraction}
After cleansing and preprocessing the dataset, we converted the app reviews in the dataset into their vector representation by using the pre-trained Bidirectional Encoder Representations from Transformers model \cite{devlin2019bert}, so-called BERT\footnote{The pre-trained BERT uncased model can be downloaded at \href{https://huggingface.co/bert-base-uncased}{https://huggingface.co/bert-base-uncased}.}. This is a language representation model trained on the BooksCorpus with 800 million words \cite{zhu2015aligning} and English Wikipedia with 2.5 billion words. The model receives a sequence of words as input and outputs a sequence of vectors. The model converted the review texts with different words into 768-dimensional vectors used as input in a machine learning model. Each of these vectors is estimated by the average of embedded vectors of its constituent words. For instance, given a review text $s$ that consists of $n$-words, $s = (w_{1}, \ldots, w_{n})$, then, $\vec{s} \approx \frac{1}{n} (\vec{w_{1}} + \ldots + \vec{w_{n}})$, where $(\vec{w_{1}} + \ldots + \vec{w_{n}})$ are the embedded vectors of $(w_{1}, \ldots, w_{n})$. Furthermore, these vectors capture both a semantic meaning and a contextualised meaning of their corresponding app reviews.


\subsubsection{Model Selection and Tuning}\label{sec:modelseletion}
Selecting a classification model that yields the optimal result is challenging. We selected five models, such as Support Vector Machine (SVM), Decision Trees (DT), Neural Network (NN), Logistic Regression (LR) and Gradient Boosting Tress (GBT) that are commonly used for text classification in the natural language processing community \cite{aggarwal2012survey}. Below is a brief description of each classification model used in our work.


    \textbf{Logistic Regression (LR)} is a linear classifier. The data is fitted into a logistic function that generates the binary output such as 0 (i.e., an honesty\_non\_violation app review) or 1 (i.e., an honesty violation app review) based on probability.
    
   \textbf{Support Vector Machine (SVM)} \cite{noble2006support} is a classifier that finds hyperplane(s) in N-dimensional space (i.e., the number of features), which can further distinguish the data into multiple categories.
   
   \textbf{Decision Trees (DT)} is one of the ensemble learners that builds trees for classification. Each tree represents a particular characteristic of the data. Given a 768-dimensional vector representation of a particular review text, DT classifies the review text into the category selected by most trees.
   
   \textbf{Gradient Boosting Trees (GBT)} is one of the ensemble learners that builds trees and boosts them for classification. When a new tree is created, it corrects errors of previous trees fitted on the same provided data. This repeatedly correcting errors process is known as the boosting process. In addition, the gradient descent algorithm is used for optimisation during the boosting process. Thus, the method is called gradient boosting trees. The model classifies app reviews into a category based on the entire ensemble of trees.
    
    \textbf{Neural Network (NN)} is a multilayer perceptron model which contains a set of interconnected layers where each layer contains a finite number of nodes. Each neural network architecture has one input layer, at least one hidden layer, and one output layer. The input data is transformed layer by layer via the activation function(s). During the training process, optimisation techniques such as stochastic gradient descent are used to optimise the performance of the model. The classified category of a particular app review is the collected result from the output layer.
    
    \textbf{Deep Neural Network (DNN)} is the extension of NN with a larger number of hidden layers that support the model to deeply learn the features in the vector representation of an app review (i.e., the embedding vector). Layers in the DNN are placed in consecutive order where the number of nodes subsequently decreases layer by layer. The first layer of the DNN is an input layer which contains $N$ number of nodes corresponding to $N$-dimensions of the embedding vector. Nodes in one layer are, then, fully connected to nodes in the next layer. The Sigmoid function is applied to transform the last hidden layer of the DNN to the output layer, which contains the classified category of an app review.

    \textbf{Generative Adversarial Networks (GANs)} \cite{NIPS2014_5ca3e9b1} is a generative neural network model that is widely used to generate high-quality data for evaluating machine learning tasks such as classification and prediction. The model consists of two networks such as the generator network and the discriminator network. The generator network learns to curate the embedding vector of an app review with an incorrect category. Both embedding vectors with the correct category and generated embedding vectors are, then, used to train the discriminator network to classify the category of app review. This aims to increase the robustness of the discriminator in classifying the category of app reviews with less amount of labelled data.


Finding the hyperparameters for models to generate optimal results is known as the fine-tuning process. We use grid search cross-validation to perform an exhaustive search to find the best set of hyperparameters for each classifier. To reproduce our results, we provide the selected hyperparameters for each selected model and the open-source GitHub repository in \cite{onlinedataset}. 

\subsubsection{Cross Validation} \label{subsec:evaluation}

To estimate the variance of the performance for each classification model, we used a 10-fold cross-validation technique. Here, we split the dataset in Section \ref{sec:dataset} into 10 chunks of data that contains an equal number of app reviews. Then, we perform the evaluation process where the training dataset contains 9 chunks of data, and another chunk of data is used as the testing dataset. Note that this is repeated until each chunk of data has been used as the testing dataset once. This approach helps us understand how well our selected models perform on unseen data.

\subsection{Results}\label{sec:resultsautomatic}
In this section, we report the results of our experiment evaluating the performance of the different machine learning models. We adopted the generally accepted metrics of \textbf{\textit{accuracy, precision, recall}}, and \textbf{\textit{F1 score}} for this purpose. Other metrics such as the Matthews Correlation Coefficient (MCC) and confusion table are shown in Table \ref{table:cm_models}. We note here that all of the models performed well (with F1 scores of 0.79 and above).

\begin{table}
\centering
\small
\caption{Comparison of the confusion matrix and Matthews correlation coefficient (MCC) of classification models.}
\label{table:cm_models}
\begin{tabular}{llllllll} 
\hline
          & SVM & LR & NN & RF & GBT & DNN & GAN \\ 
\hline
True negative  & 0.432  &  0.407  &  0.358  &  0.371     & 0.358 & 0.407 & 0.383  \\
True positive &  0.457  & 0.469 & 0.482  &  0.420 &  0.420 & 0.506 & 0.482 \\
False positive &  0.025  &  0.049  &  0.099  &   0.085     &        0.099  & 0.049 & 0.074      \\
False negative  &   0.086    &  0.074  & 0.062  &     0.124    &  0.124 & 0.037 & 0.062 \\
MCC & 0.785 &  0.753 & 0.676 & 0.581 & 0.555 & 0.826 & 0.726 \\
\hline
\end{tabular}
\end{table}


Table \ref{table:models} shows the results of our 7 different machine learning classification algorithms. \textbf{The DNN algorithm came out to be the best performing model with an accuracy of 0.914, precision of 0.911, recall of 0.932, and an F1 score of 0.921}. The second-best performing algorithm is the SVM model, with an accuracy of 0.889, precision of 0.949, recall of 0.841, and an F1 score of 0.892. The high performance of our DNN model makes it useful in practical applications for detecting the violation of the value of honesty in reviews.


\begin{table}
\centering
\small
\caption{Comparison of classification models.}
\label{table:models}
\begin{tabular}{llllllll} 
\hline
          & SVM & LR & NN & RF & GBT & DNN & GAN  \\ 
\hline
Accuracy  & 0.889                  & 0.877               & 0.840          & 0.790         & 0.778   & 0.914  & 0.864          \\
Precision & 0.949                  & 0.905               & 0.830          & 0.829         & 0.810  & 0.911 & 0.867            \\
Recall    & 0.841                  & 0.864               & 0.886          & 0.773         & 0.773  & 0.932 & 0.886            \\
F1 score  & 0.892                  & 0.884               & 0.857          & 0.800         & 0.791  & 0.921 & 0.876           \\
\hline
\end{tabular}
\end{table}

One of the aims of our work is to introduce an automatic method for detecting honesty violation reviews that performs better than current approaches. Similar studies on text classification have compared their approaches to either the current state-of-the-art or a baseline random classifier \cite{AlOmar:2021,Maldonado:2017}. Hence we compare our best-performing machine learning model (DNN) with a baseline random classifier only since there is no current state-of-the-art in detecting the violation of honesty in app reviews, similar to what recent works have done \cite{AlOmar:2021,Maldonado:2017}.




We used the statistics of our dataset to compute the metrics of the random classifier. The precision of a random classifier can be computed by dividing the number of honesty violation reviews by the total number of reviews: \[ precision = \frac{401}{236,660} = 0.0017\] 

The recall is $0.5$, as there are only two outcomes for a review classification: honesty violations reviews or honesty\_non\_violations reviews, with a $0.5$ probability of a review containing the violation of the value of honesty. Based on the precision and recall values, we compute the F1 score of the baseline random classifier as:

\[F1 \ score = 2 * \frac{0.0017 * 0.5}{0.0017 + 0.5} = 0.0034\]

Table \ref{table:compare} summarises the comparison of our best-performing machine learning model (DNN) with the baseline. As can be seen, the DNN model has a better performance than the  baseline random classifier. Our DNN model has an F1 score of $0.921$, while the baseline random classifier has an F1 score of $0.0034$, respectively. Table \ref{table:compare} also shows that our DNN model surpasses the baseline random classifier by 270.882 times in detecting honesty violation reviews.


\begin{tcolorbox}[boxrule=1pt, colback=gray!10!white, boxsep=0pt, top=7pt, bottom=7pt, left=7pt, right=7pt, before skip=3pt,after skip=3pt]
\small
\textit{\textbf{RQ1 Answer}:} The DNN model surpasses the baseline random classifier in identifying the violation of the value of honesty in reviews. Our model achieves an F1 score of 0.921 with an improvement of 270.882 times the baseline random classifier in classifying honesty violation reviews from honesty\_non\_violation reviews.
\end{tcolorbox}


\begin{table}
\centering
\small
\caption{Comparison of our model to a baseline classifier.}
\label{table:compare}
\begin{tabular}{l|lll|lll} 
\hline
               & \multicolumn{3}{l|}{Our (DNN) approach} & \multicolumn{3}{l}{Random classifier}  \\
               & Precision & Recall & F1           & Precision & Recall & F1                \\ 
\hline
Classification & 0.911         & 0.932      & 0.921            & 0.0017    & 0.5    & 0.0034            \\ 
\hline
Improvement    & -         & -      & -            & 535.882x         & 1.864x      & 270.882x                 \\
\hline
\end{tabular}
\end{table}
\begin{table}[]
\caption{Frequency $(f)$ of app reviews in the honesty violation categories (out of 401 total \textbf{\textit{honesty\_violations}} reviews -- note that some reviews fall into multiple categories).}
\label{table:categories}\centering
\resizebox{0.6\textwidth}{!}{%
\begin{tabular}{@{}ll@{}}
\toprule
\textbf{Honesty Violation}              & \textbf{\textit{f}}   \\ \midrule
Unfair cancellation and refund policies & 48 (12\%)    \\
False advertisements                    & 55 (14\%)    \\
Delusive subscriptions                  & 33 (8\%)     \\
Cheating systems                        & 93 (23\%)    \\
Inaccurate information                  & 15 (4\%)     \\
Unfair fees                             & 106\% (26\%) \\
No service                              & 64 (16\%)    \\
Deletion of reviews                     & 6 (1.5\%)    \\
Impersonation                           & 9 (2\%)      \\
Fraudulent-looking apps                 & 29 (7\%)     \\ \bottomrule
\end{tabular}%
}
\end{table}

\section{Categories of Honesty Violations (RQ2)}\label{sec:categorieshonesty}
\subsection{Categorisation Approach}\label{sec:approachRQ2} While the machine learning models in Section \ref{sec:automaticclassification} could effectively distinguish between honesty violations reviews and honesty non-violations reviews, we are also interested in understanding the types of honesty violations reported in reviews. To this end, we applied the open coding procedure \cite{glaser1968discovery} on the 401 \textit{\textbf{honesty}\textbf{\_}\textbf{violations}} reviews. As discussed in Section \ref{sec:dataset}, these reviews include honesty violations. First, an analyst followed the open coding technique to label all these 401 reviews and identified 10 types of honesty violations. The 401
\textit{\textbf{honesty}\textbf{\_}\textbf{violations}} reviews were assigned to these 10 categories. The results of the open coding were stored in an Excel spreadsheet file and shared with the second and third analysts. 
Then, the second analyst cross-checked the first 100 labelled reviews while the third analyst cross-checked the remaining 301 labelled reviews. Next, the first analyst held Zoom meetings with the second and third analysts to discuss and resolve the conflicts and disagreements. Note that the disagreements were resolved using the negotiated agreement approach \cite{2013_campbell_coding,1974_morrissey_sources}.

\subsection{Results}\label{sec:resultsRQ3}

Our analysis of the $401$ \textit{\textbf{honesty}\textbf{\_}\textbf{violations}} reviews revealed $10$ categories of honesty violations reported in app reviews. Below we provide a definition of these categories, sample reviews, and a summary of their prevalence. While we highlight the different categories within the violation of the value of honesty and provide example reviews, we note that the categories are not mutually exclusive. Table \ref{table:categories} shows these categories and the frequency of the corresponding reviews per category.


\subsubsection{Unfair cancellation and refund policies} This category covers all reviews where the users perceive the cancellation and refund policy as unfair, nontransparent, or deliberately misleading. It also includes situations where the user feels that the developers deliberately make it difficult for the user to cancel their subscription. For example, in some apps, the user can sign up for a subscription with the click of a button within the app but cannot cancel the subscription from within the app; the user is asked to log in to a website to cancel the subscription. In other cases, the cancellation instruction is not clear and leads to a loop of cancellation steps. Examples of reviews claiming these practices include:

\begin{myquote}
\noindent \faThumbsODown \hspace{0cm} 
\textit{“The app allows you to accidentally sign up to premium with a push of a button. When you want to cancel, however, you can't do that via the app... You have to go to the webpage, enter details and cancel there.”} \end{myquote}

\begin{myquote}
\noindent \faThumbsODown \hspace{0cm} 
\textit{“Deceptive billing practices - information on cancelling is circular; emailed a link that advises to email. [It] doesn't have colour tag functionality across web and app; very poor UX and worse customer service.”} \end{myquote}

Sometimes, the app also makes it easy for the user to mistakenly activate a premium subscription in the way the interface and flow are designed, e.g.:

\begin{myquote}
\noindent \faThumbsODown \hspace{0cm} 
\textit{“Use with caution. It's unscrupulous about signing you up for a subscription when you're skipping past the in-app ads. It's not made clear once you've subscribed, and there's no way of cancelling it through the app.”} \end{myquote}

Another aspect of this category focuses on situations where the user perceives the refund steps and policies to be dishonest and unfair. This also involves situations where the refund policy does not cater to accidental subscriptions, e.g.:

\begin{myquote}
\noindent \faThumbsODown \hspace{0cm} 
\textit{“DO NOT SIGN UP FOR FREE TRIAL! IT IS A SCAM. YOU WILL GET CHARGED ANYWAY, AND YOU WILL NEVER GET YOUR MONEY BACK!! Once again, after numerous attempts to blame Google, this developer has still not refunded my \$38. Once again, I cancelled 3 full days before the free trial ended but was still charged. Once again, [I] contacted the developer, who told me that I would receive a full refund within 7 to 10 days, and still nothing. I have saved the email, pricing this to be true. DO NOT TRUST THIS DEVELOPER. SCAM!!!!”} \end{myquote}


\subsubsection{False advertisements} This category relates to situations where the user perceives that the advertised features and functionalities of the app as described by the developers are not contained in the app. The user downloads the app or pays for a subscription on the basis of accessing certain functionalities or features only to find out the descriptions, including screenshots on the app distribution platform are different from the actual functionalities available in the app. Two examples of these are shown below:

\begin{myquote}
\noindent \faThumbsODown \hspace{0cm} 
\textit{“Couldn't find Google Assistant integration anywhere. Even though it's been advertised everywhere when searching the web for the app... It's even in the description of the app here. That's false advertising. I will edit my review when it's out of Beta and working in the final version.”}  \end{myquote}

\begin{myquote}
\noindent \faThumbsODown \hspace{0cm} 
\textit{“The app doesn't listen to the watch at all. I've tried completing and snoozing and it does nothing. The watch app can only add tasks, so the screenshots they're sharing here are DECEPTIVE.”} \end{myquote}

In some cases, the app lures users into downloading the app on the basis that it is free-for-use only for the user to find out that the free-for-use is a trial version for a specific time period and not perpetually free as implied in the app description:

\begin{myquote}
\noindent \faThumbsODown \hspace{0cm} 
\textit{“The actual free version doesn't allow you anything, not even to learn how to use the app properly. That role is filled by 7 days of free premium. The free, on the description, is a lie. Is a paid-only app with temporary free access to its full features that gets practically useless after the 7-day trial... I don't like to be lied to.”} \end{myquote}

In addition, the app developers (through the app description) make promises to users to give them certain benefits like a free premium subscription when a particular action is carried out (e.g., inviting a particular number of friends to sign up). However, they never truly fulfil their promises when the user fulfils their end of the bargain. These unfulfilled obligations are perceived by the end-user as a violation of honesty, e.g.: 

\begin{myquote}
\noindent \faThumbsODown \hspace{0cm} 
\textit{“I love this app however I sent the link to several friends and they got the app and I received no premium time whatsoever. Don't be dishonest with your apps. That's lame.”} \end{myquote}

Another example relates to scenarios where the user is invited to make certain commitments based on a future reward and the developers bail out on their prior commitment: 

\begin{myquote}
\noindent \faThumbsODown \hspace{0cm} 
\textit{“Shame on Them! Liars. I paid for the season pass TWICE (ONCE for my apple device and the other for my Samsung Device). I was falsely promised access to ALL FUTURE CONTENT. Now they are trying to charge me for the Parisian Inspired TOKENS! HOW DARE THEY LIE AND BAIT AND SWITCH.”} \end{myquote}

\subsubsection{Delusive Subscriptions} Any review describing complaints related to unfair or nontransparent automatic subscription processes are classified under this category. There are instances where no notifications are provided to let the user know they are subscribed to the app or premium version of the app, and the user only finds out about the subscription from the deductions in their bank accounts:

\begin{myquote}
\noindent \faThumbsODown \hspace{0cm} 
\textit{“I just realised that I have been charged for some crappy premium service fee which I had no idea about when using the app. Why is this charge by default? Why was I not informed in the first place? Beware of scam for useless monthly premium fees!”} \end{myquote}

\begin{myquote}
\noindent \faThumbsODown \hspace{0cm} 
\textit{“I can't believe I was charged 55.99. What are you giving me? Gold? I unsubscribed but saw mysterious charge in my bank account.”} \end{myquote}

Additionally, there is the issue of lack of user consent in the subscription process where certain apps do not provide a confirmation mechanism that prevents accidental subscriptions by the user, e.g.:

\begin{myquote}
\noindent \faThumbsODown \hspace{0cm} 
\textit{“Made me pay 1 year worth of subscription without my confirmation. Only used its free trial because I had to use it once. What a scam...”} \end{myquote}

In some scenarios, the automatic subscription is hidden behind an in-app ad/feature, and an unsuspecting user who clicks on the feature is automatically subscribed to the premium version of the app without a clear warning or confirmation, e.g.:

\begin{myquote}
\noindent \faThumbsODown \hspace{0cm} 
\textit{“Deceptive practices. If you click the in-app "ad" that simply says enable notifications, you'll automatically be signed up and billed for their premium service. This bypasses the Google/Apple stores subscription model and bills your card directly. Not to mention it's impossible to downgrade from this service in the app itself; you have to visit their website, which is a deliberately obstructive hurdle considering you can upgrade in the app just fine.”} \end{myquote}


\subsubsection{Cheating systems} All reviews concerning the user’s perception of fraud by other persons or cheating within the inner workings of the app are classified under this category. Users complain of unfairness in either the process or outcome of the app, especially processes/outcomes that are supposedly statistically random. While accusations of this kind from the users are prevalent and subjective, they may not really be the case. However, we labelled these kinds of reviews based on the \emph{perception} of the users as captured in their comments. Reviews related to this category are mostly found in games or game-like systems. For example:

\begin{myquote}
\noindent \faThumbsODown \hspace{0cm} 
\textit{“This game cheats. It uses words not found in the dictionary. Also it told me a word was unplayable, but it was the first best word option.”} \end{myquote}

\begin{myquote}
\noindent \faThumbsODown \hspace{0cm} 
\textit{“I play it with my sister often. However, there is the problem of the game and AI cheating. I rolled a 2 and a 3 at the start of the game and it moved me FOUR spaces forward not five. Four. That happened several times and I can assure you I was looking everytime it happened. I am very disappointed at the fact this game is cheating...”} \end{myquote}

In some of the reviews, users complain that the game works properly when the user loses and parts with money and only freezes when the AI system in the app is about to lose. Based on the reviews, the users seem to be using real money in the games/apps. This complaint is a recurring theme within this category:

\begin{myquote}
\noindent \faThumbsODown \hspace{0cm} 
\textit{“You have to pay for it, then the game just freezes when you win against the CPU? Reset it over and again, keeps freezing unless it rolls something to not land on my property. Also, is the dice rigged against the CPU? Honesty? With as much as I owned in the beginning, none of the 3 CPUs would land on anything I owned. Anytime the last CPU needs to raise money, game freezes, guess ya just can't win.”} \end{myquote}

\begin{myquote}
\noindent \faThumbsODown \hspace{0cm} 
\textit{“there's a glitch in it that freezes the game from continuing when you're winning. The dice just disappears, but the trains and clouds and aircrafts keep moving. It's like It is designed so that one doesn't win them.”} \end{myquote}

\begin{myquote}
\noindent \faThumbsODown \hspace{0cm} 
\textit{“When playing against the computers when you're about to win and bankrupt the final computer the game conveniently freezes. It does not allow you to win. Not a very fun game to play, I want my money back.”} \end{myquote}

We consider this category important as some of these apps require the use of real money to play or for in-app purchases. If apps are dishonest in the underlying process of the systems that are expected to be fair, then that constitutes not only a violation of the value of honesty, it might potentially be a crime. This is worth considering, especially when the exact issue is raised by several users: 

\begin{myquote}
\noindent \faThumbsODown \hspace{0cm} 
\textit{“Although you say that the dice is random, i cannot help but feel that it is rigged. Take a look at your reviews, there are many other players that feel the same. Can't be all of us are wrong. Or maybe we are suffering from mass hysteria?”} \end{myquote}

Other non-game examples include cases where the user reports not having the full value of the fee they were charged for the app and feels cheated. For instance:

\begin{myquote}
\noindent \faThumbsODown \hspace{0cm} 
\textit{“Whenever I pay for parking the app always steals 5 minutes off my parking time. For example, I pay for 60 minutes and the timer starts at 54 minutes and 59 seconds. I am very upset, this has been happening for a while and probably to many more people as well. That is a lot of money!”} \end{myquote}

\begin{myquote}
\noindent \faThumbsODown \hspace{0cm} 
\textit{“This app will not give you re requested amount of parking time. If you park for 15 minutes it will immediately say you have ~11 minutes left. I understand that you have to charge but at least give me the requested amount of parking time.”} \end{myquote}

\subsubsection{Inaccurate information}
This category covers where users perceive that the app provides false or inaccurate information as captured in their reviews. This includes situations where inaccurate information can increase the likelihood of the user inadvertently making wrong selections at a cost to them. In the review below, the user complains the design of an app feature tricks them into paying for the wrong parking spot:

\begin{myquote}
\noindent \faThumbsODown \hspace{0cm} 
\textit{“When you need to pay for additional time, and click 'Recent' to pay for the most Recently parked in place - the first item is not the place you just parked in so it tricks you into paying for the wrong place (dark pattern). Please make the Recent accurately reflect the most recently parked in place.”} \end{myquote}

Another example review in this category is quite severe as it relates to a health emergency app providing potentially inaccurate information that might be detrimental to the user: 

\begin{myquote}
\noindent \faThumbsODown \hspace{0cm} 
\textit{“Try to use this in an actual emergency and you'll just end up as a dead idiot holding a cellphone. The information is either useless or completely false in most cases. Don't bother downloading.”} \end{myquote}

Other less severe but important reviews where the user perceives the app provides inaccurate information or notification are shown below: 

\begin{myquote}
\noindent \faThumbsODown \hspace{0cm} 
\textit{“Do not buy unless you are sure you want to. You will NOT be able to get it set up and working within the 15 minute refund window. The instructions online are so cryptic it (and wrong).”} \end{myquote}

\begin{myquote}
\noindent \faThumbsODown \hspace{0cm} 
\textit{“Very annoying every time when you open the app it shows you have a notification. Then checking your notifications you don't have any.”} \end{myquote}

\subsubsection{Unfair fees}
This category relates to issues surrounding what the user considers to be unfair fees or charges. This also applies to cases where the user feels that they have not received a fair deal or that the app charges more money than it ought to. Because the definition of honesty also covers fairness, we also consider these kinds of issues a potential violation of the value of honesty. In the example below, the user complains of being charged more than they think is fair; they were charged a car parking rate for parking a bike.

\begin{myquote}
\noindent \faThumbsODown \hspace{0cm} 
\textit{“Went through the sign up process and parked my bike in a bike parking zone. Put in the correct zone details for the bike parking area and got charged a car parking rate. Rang support and they said there is no bike parking at that location. I explained there was and they told me to ring the council.”} \end{myquote}

Other examples of fees considered by the user to be unfair are:

\begin{myquote}
\noindent \faThumbsODown \hspace{0cm} 
\textit{“The app charges you 0.25 per transaction. So I paid 0.75 to pay for parking it charged me 0.25 service fee then I extended my parking 0.25 and it charged me again 0.25!!! Biggest scam in the world.”
} \end{myquote}

\begin{myquote}
\noindent \faThumbsODown \hspace{0cm} 
\textit{“The only annoying things are that I have to buy any extra Monopoly Board in the same game when I already paid the main game. Can you not give extra Monopoly Boards in the same game for free. You are not fair!”}  \end{myquote}

This category is also reflected in the form of hidden charges where the user is not aware of subsequent charges made to their account. These hidden charges can take the form of a vague bill (as shown in the review below) or not notifying the user with respect to extra charges.

\begin{myquote}
\noindent \faThumbsODown \hspace{0cm}
\textit{“This is a notorious company with horrible app I've ever used. They hide the history and details very deep for you to check and trace. And the monthly bill is also vague. I experienced they secretly bill me!”}  \end{myquote}

\begin{myquote}
\noindent \faThumbsODown \hspace{0cm}
\textit{“LOOK OUT PEOPLE. THIS IS A SCAM. THEY DID NOT WARN OF A DEPOSIT FEE AND THEY TOOK 33\% OF THE DEPOSIT. I RECOMMEND SUING THEM NOW.”} \end{myquote}

Another related issue within this category is dubious charges where the user account has been charged, and it is not clear why those charges occur. Abnormally high fees (more than the standard subscription fees) and overcharging of the user account are also captured under this category. For example:


\begin{myquote}
\noindent \faThumbsODown \hspace{0cm}
\textit{“It charged me £74.50 when I bought a ticket for £1.50 it's a absolute scam I want my money back!” }\end{myquote}


\subsubsection{No service}
This category mainly covers reviews in which the user complains of not being able to access the app's main functionality after purchase, leading to undesirable consequences for the user. The main difference between the \textit{false advertisement} category and this category is that the former deals with features/functionalities of the app that do not work as advertised. The latter deals with situations where the app does not work at all, i.e., does not even serve its main purpose for the user after the user has made financial commitments in the form of a purchase or subscription. In the example below, the user is fined for illegal parking after paying for parking using the app:

\begin{myquote}
\noindent \faThumbsODown \hspace{0cm}
\textit{“Horrible experience with this app. Causing a lot of frustrations with users. when it fails and I get a ticket there is no much help I can get. sometimes I just pay the fines just because the complaint system is awfully inconvenient. I feel cheated and it looks like a money making tool for whoever is collecting the fines.”
} \end{myquote}

Another related example is shown below:

\begin{myquote}
\noindent \faThumbsODown \hspace{0cm}
\textit{“I spent 20 euros with all the DLCs included, I feel pretty deceived not being able to play the game.”}\end{myquote}



\subsubsection{Deletion of reviews}
This category highlights reviews where the app developers are suspected of deleting reviews left by the user, especially negative reviews. A review captures user feedback, describing their experience of an app, and intending users of an app typically consult the reviews left by other users on the app distribution platform before downloading the app \cite{Obie:2020}. Thus, the act of deleting unfavourable reviews by the app developers is perceived as a dishonest practice by the users because leaving only positive reviews may not paint an accurate picture of the app. Users may also feel like the app developers are trying to hide their complaints or other nefarious practices. 

It can be argued that certain comments are deleted by app developers because those comments contain ad hominem attacks from the users instead of complaints relating to the app itself. While it is debatable whether app developers are justified in deleting perhaps vitriolic ad hominem comments, we do not make any judgement as to this but simply categorise users' perceptions and complaints of this practice as captured in their reviews. Examples of reviews depicting this accusation are shown below:

\begin{myquote}
\noindent \faThumbsODown \hspace{0cm}
\textit{“I left them a negative review and the developer deleted it. Now I'm going to review them on YouTube and all social media platforms. Basically, they are scammers.”} \end{myquote}

\begin{myquote}
\noindent \faThumbsODown \hspace{0cm}
\textit{“Deleted my honest review. Warning. Steer clear. They keep trying to make you slip up and pay for premium. I signed up for a free trial last year and they make it too difficult for you to find where to cancel. Was charged about \$40...  shame such a good app is tarnished by such shady practices.”} \end{myquote}

\subsubsection{Impersonation}
An impersonation is an act of pretending to be another person or entity \cite{coldictionary:2021}. It also involves the act of giving a false account of the nature of something. This category covers all reviews relating to impersonation or misrepresentation by the app or app developers. This includes scenarios where an app pretends to have the authority of (or relationship to) an organisation when in reality, it has no such relationship. An example review is captured below:

\begin{myquote}
\noindent \faThumbsODown \hspace{0cm}
\textit{“STAY AWAY... this app is a scam. the stickers make it look like it’s Brisbane council approved. it’s not and they are no help. I still got a fine for using the app correctly and the Brisbane council parking police have no access to check if you have paid or not and do not accept this as a payment method.”} \end{myquote}

Another example in this category reflects situations where users feel that they are interacting with bots instead of humans when they have signed up to the platform to interact with humans. This is similar to false advertising-related lawsuits of the Match.com platform described in section \ref{sec:introduction}. An example of this is:

\begin{myquote}
\noindent \faThumbsODown \hspace{0cm}
\textit{“Good game, fake players online. I wanted a challenging Monopoly game. But when I start. I can tell that some are bots not real people online. For example, they quickly trade when it is their turn. A normal human will take some time to choose options.”} \end{myquote}


\subsubsection{Fraudulent-looking apps}
This category includes reviews reporting suspicious-looking apps based on observations of users or apps deemed to be fake by the users. We created a separate category for these kinds of reviews. Although the users flag the apps in these reviews as fraudulent, they do not provide specific reasons for their accusations beyond their perception of the app as fake or fraudulent. Furthermore, these types of reviews do not fit any of the categories described above, and we sought to highlight them based on the user accusations captured in their reviews. Examples of these reviews include: 


\begin{myquote}
\noindent \faThumbsODown \hspace{0cm}
\textit{“...Be careful with this kind of dishonest apps”}\end{myquote}

\begin{myquote}
\noindent \faThumbsODown \hspace{0cm}
\textit{“This is a fraud app don't download”} \end{myquote}

\begin{tcolorbox}[boxrule=1pt, colback=gray!10!white, boxsep=0pt, top=3pt, bottom=3pt, left=3pt, right=3pt, before skip=3pt,after skip=3pt]
\small
\textbf{\textit{RQ2 Answer:}} The result of our analysis of the honesty violations dataset shows that honesty violations can be characterised into ten categories: unfair cancellation and refund policies, false advertisements, delusive subscriptions, cheating systems, inaccurate information, unfair fees, no service, deletion of reviews, impersonation, and fraudulent-looking apps. 
\end{tcolorbox}
\section{Developer experience with honesty violations in mobile apps (RQ3)}\label{sec:RQ3}

\subsection{Practitioner Study Design Approach}

Our analysis of app reviews in RQ1 and RQ2 indicates that honesty violations exist in mobile apps from the perspective of end users. But what about app developers' experience with them? This motivated us to explore mobile app developers' experience with honesty violations in mobile apps.

\begin{figure}
    \centering
    \includegraphics[width=\textwidth, height=5cm]{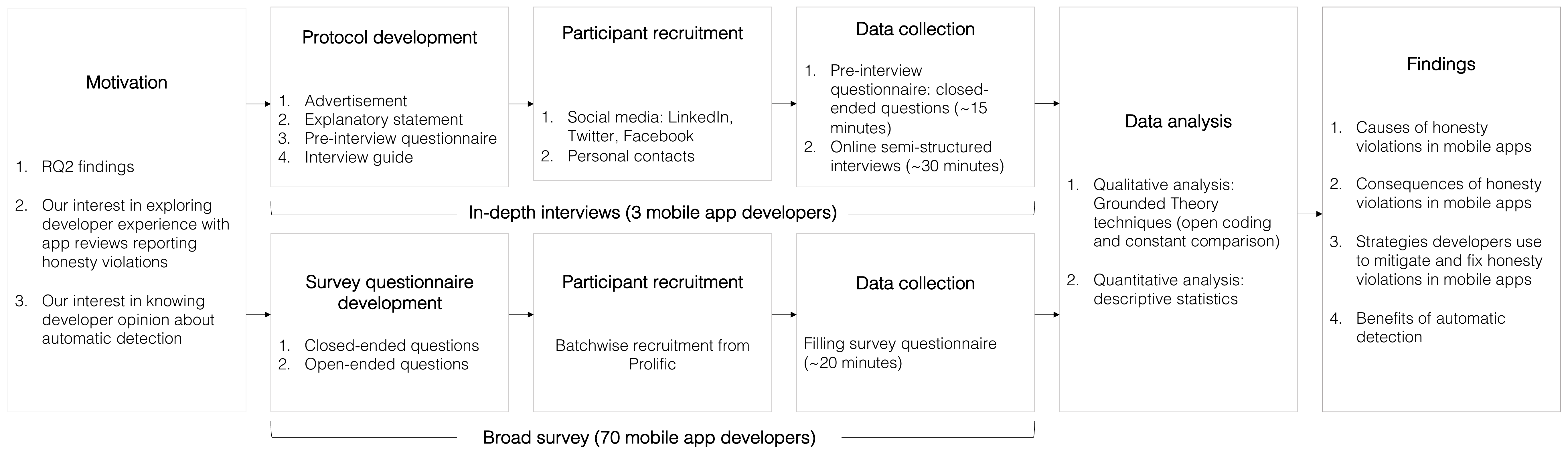}
    \caption{Developer Study}
    \label{fig:dx_method}
\end{figure}

We took an interview and survey-based approach, referred to as the developer study, (Fig. \ref{fig:dx_method}) to understand developers' experience with honesty violations in the mobile apps they develop. In parallel, we conducted a set of in-depth semi-structured interviews, and we conducted a broad survey -- both with mobile app developers. Collecting data from both interviews and surveys strengthened our findings well. The replication package, which consists of the artefacts we developed to collect data in both studies, is available online\footnote{https://github.com/kashumi-m/ReplicationPackageMobileAppsHonestyViolations}. In this section, first, we explain the interview study and then the survey study.

\subsubsection{Step Int: Interview Study}

\textbf{Step Int.1: Protocol Development: }
For the interview study, to recruit participants, we prepared an advertisement and an explanatory statement; and to collect data, we prepared a pre-interview questionnaire and an interview guide. The details about the artefacts are explained under participant recruitment and data collection below.

\noindent \textbf{Step Int.2: Participant Recruitment: }
We recruited participants by sharing the explanatory statement and posting an advertisement on social media such as LinkedIn, Twitter, and Facebook with a link to the explanatory statement. The explanatory statement consisted of details of the study, including the procedure, potential benefits, and how we preserve the confidentiality of the participants. Potential participants contacted us, showing their interest in participating in the interview study. We recruited three participants to proceed with the data collection.

\noindent \textbf{Step Int.3: Data Collection: }
Data collection for the interview study consisted of two parts: a pre-interview questionnaire and an online interview. 

\textit{Pre-interview Questionnaire.} Each participant was given a pre-interview questionnaire to fill in before the interview. The questionnaire consisted of questions about their demographics (age, gender, country of residence, professional experience, including total experience in the software industry and total mobile app development experience), context (type of apps the participants develop based on the list of app types as in Google Play Store, types and frequency of honesty violations the participant had experienced for their apps, who is responsible for honesty violations in mobile apps, and the participant's opinion about automatic detection of honesty violations in mobile apps (usefulness, beneficiaries, how beneficial). We included the definition of honesty and honesty violations at the beginning of the pre-interview questionnaire so that participants' interpretation of the terms aligns with ours. We also repeated the definition of honesty violations at the beginning of each relevant question section to ensure that participants' interpretation of the term remains the same until the end of the questionnaire. The pre-interview questionnaire was hosted on Qualtrics\footnote{https://www.qualtrics.com/} and took around fifteen minutes to complete. Having a pre-interview questionnaire helped us in collecting data on closed-ended questions early so that we had a high-level understanding of participants' background with honesty violations in mobile apps and also led us to have more time for in-depth discussions during the interviews.

\textit{Online Interview.} After we confirmed that participants had filled out our pre-interview questionnaire, we conducted  online interviews with them at an agreed time using Zoom. Each semi-structured interview lasted approximately thirty minutes and was audio recorded. During the interviews, we focused on asking open-ended questions from the participants so that we could gain much more rich data on their experiences with honesty violations in mobile apps. The interviews started by showing some examples of honesty violations reported in mobile app reviews to the participants. This made it easy for the participants to answer the questions. First, we asked the participants about the reasons for honesty violations happening in mobile apps, then asked about the impact of honesty violations on end users and developers/owners of the mobile apps. After that, we asked participants what possible strategies exist to avoid honesty violations. We also asked them what strategies they adopted to address honesty violations that they had encountered in their mobile apps (if any). The interviews ended with an open question, allowing the participants to share anything else they liked to share about honesty violations in mobile apps. After the interviews ended, the audio recordings were transcribed using Otter\footnote{https://otter.ai/}.

\begin{figure}
    \centering
    \includegraphics[width=\textwidth,height=7cm]{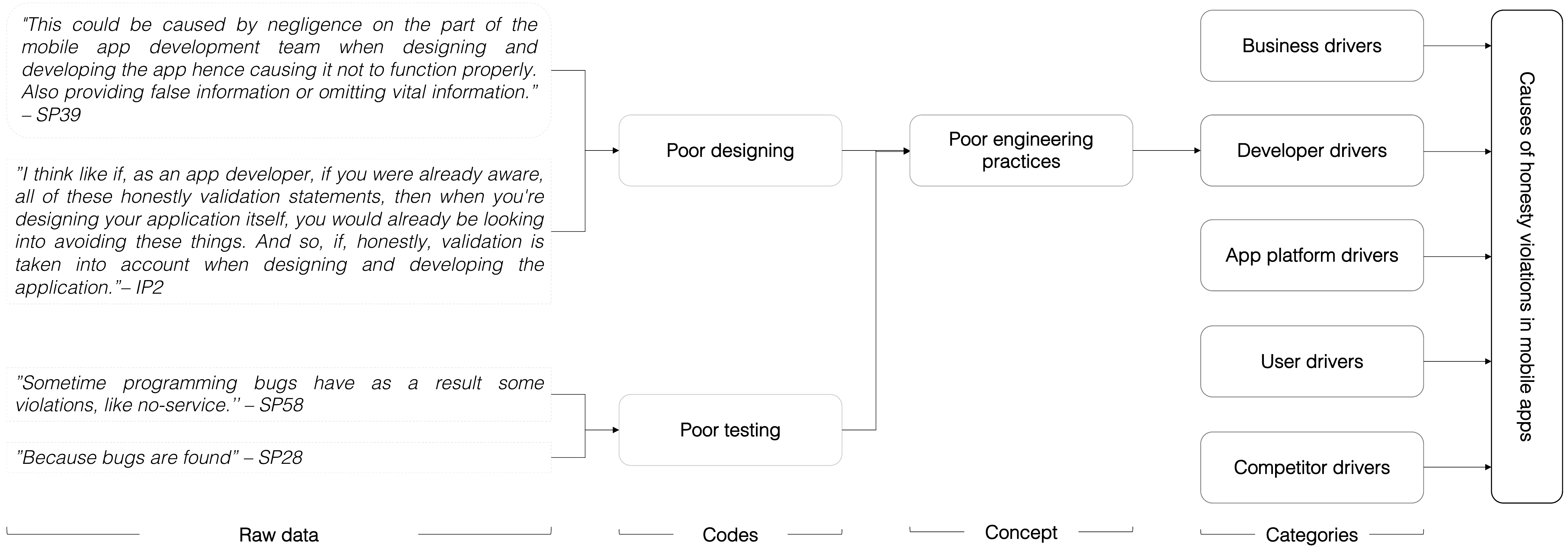}
    \caption{Example of Qualitative Analysis: Investigating Causes of Honesty Violations in Mobile Apps}
    \label{fig:dx_qual_example}
\end{figure}

\textbf{Step Int.4: Data Analysis:}
\textit{Qualitative data analysis.} The qualitative data collected were analysed using open coding and constant comparison techniques. We used MAXQDA\footnote{https://www.maxqda.com/} to analyse the data. The responses to the questions (raw data) were interpreted in small chunks of words (codes), and they were constantly compared to group similar codes together to develop the concepts. The concepts were then constantly compared to develop categories. Fig. \ref{fig:dx_qual_example} shows an example of qualitative analysis. 
\textit{Quantitative data analysis.} As some closed-ended questions of the interview study were repeated in the survey study, the quantitative data were analysed together with the quantitative data collected from the survey study. Descriptive statistics were used to analyse the data.

\subsubsection{Step Survey: Survey Study}

\textbf{Step Survey.1: Survey Questionnaire Development: }
We developed a questionnaire with a mix of open-ended and closed-ended questions. The survey contained demographics, contexts, causes, consequences, strategies, and automatic detection of honesty violations in mobile apps. The questions on demographics, context, and automatic detection were the same questions we used in the pre-interview questionnaire of the interview study. We further used open-ended questions to allow participants to freely share their experiences about causes, consequences, and strategies.

\textbf{Step Survey.2: Participant Recruitment: }
We used Prolific\footnote{https://www.prolific.co/} to recruit participants for the survey study. The participants were recruited batch-wise, i.e., 10 x 5, which altogether resulted in recruiting 70 participants.

\textbf{Step Survey.3: Data Collection: }
The data collection of the survey study was straightforward. The link to the survey questionnaire hosted on Qualtrics was shared through a Prolific post. The participants took around twenty minutes on average to complete the survey.

\textbf{Step Survey.4: Data Analysis: }
The same procedure as in the interview study was followed to analyse the collected data.

\subsection{Interview and Survey Study Results: Participant Information and Their Context}

\subsubsection{Participant Information.}
\begin{figure}
    \centering
    \includegraphics[width=\textwidth]{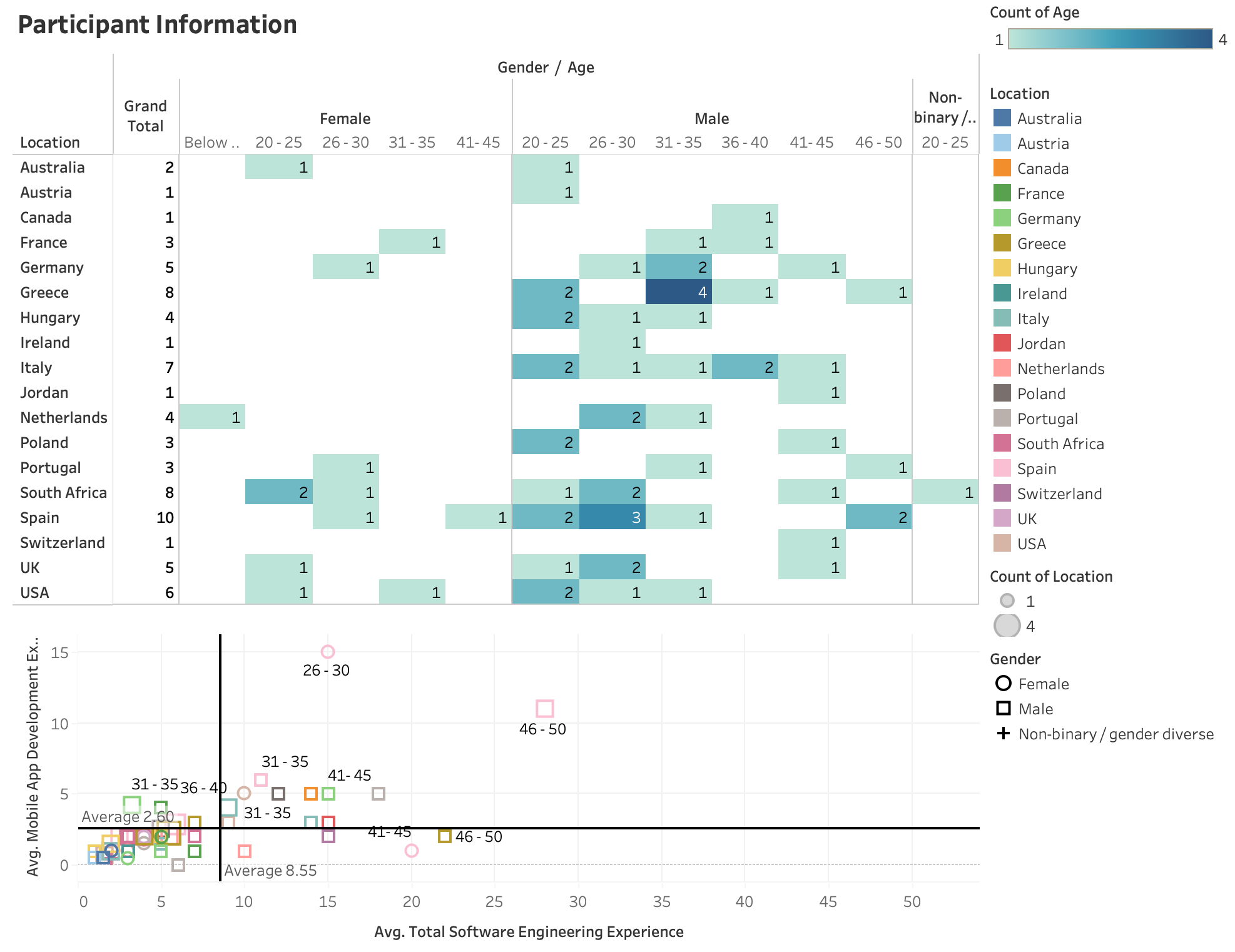}
    \caption{Participant Information (Excluded in Figure: One participant had mentioned total work experience as 82 years but within the age range of 31-35; Included in Figure: One participant with less than 1 year of mobile app or without mobile app development experience)}
    \label{fig:dx_participant_info}
\end{figure}

Fig. \ref{fig:dx_participant_info} is a summary of the participant information (location, gender, age, total experience, mobile app development experience). The majority of the participants were from Spain (10 participants), followed by South Africa and Greece (8 participants each); were male (59 participants); had an average total software engineering experience of 8.55 years, and an average mobile app development experience of 2.60 years.

\subsubsection{Types of Mobile Apps Participants Develop.}
\begin{figure}
    \centering
    \includegraphics[width=\textwidth]{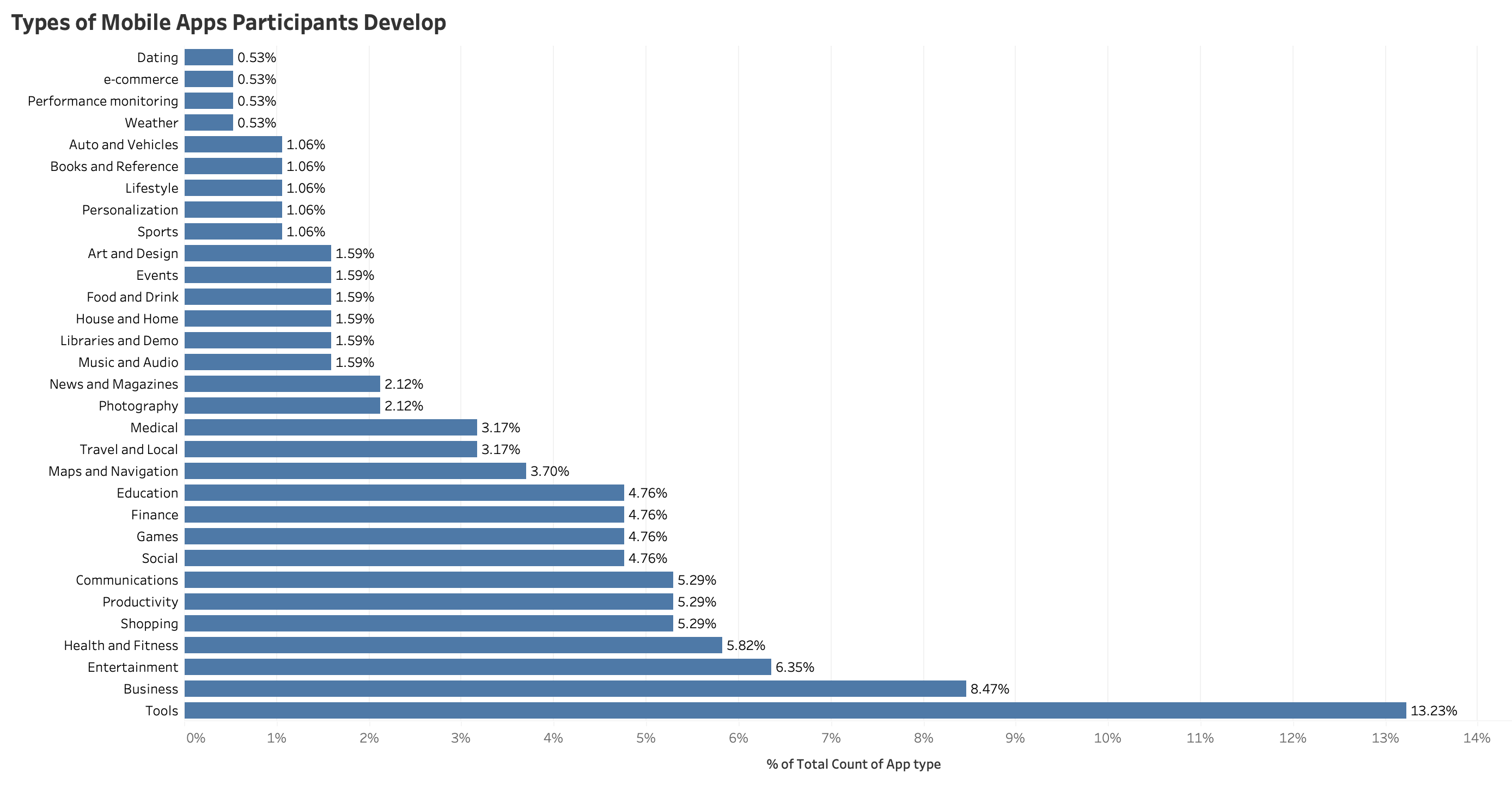}
    \caption{Types of Mobile Apps Participants Develop}
    \label{fig:app_types}
\end{figure}

A summary of the types of mobile apps our participants develop is shown in Fig. \ref{fig:app_types}. While the developers are not limited to developing one type of app, our participants selected many types, and the key app type they mentioned as they develop is \textit{tools} (13.33\%).

\subsubsection{Developer Experience: Reported Honesty Violations in App Reviews.}
\begin{figure}
    \centering
    \includegraphics[width=\textwidth]{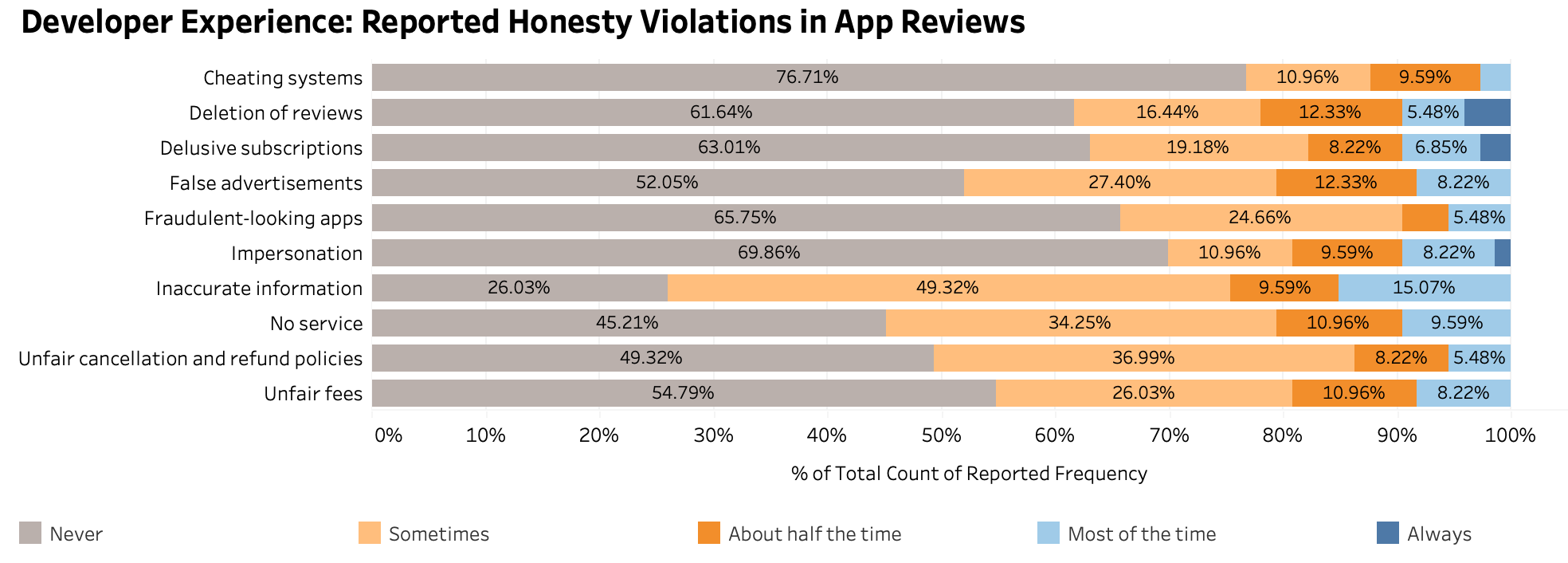}
    \caption{Developer Experience in Receiving Honesty Violations in Mobile App Reviews}
    \label{fig:dx_hc_counts}
\end{figure}

We asked our participants which types of honesty violations they received for their apps. The results are shown in Fig. \ref{fig:dx_hc_counts}. According to our participants' experience, the most reported honesty violation by the users is \textit{inaccurate information} (sometimes+about half the time+most of the time=73.97\% of participants). This is followed by \textit{no service} (sometimes+about half the time+most of the time=54.79\% of participants).

\subsection{Interview and Survey Study Results} 
We found several causes, consequences, and mitigation and fixing strategies for honesty violations in mobile apps. Further, we found how useful automatic detection of honesty violations is, the benefits of automatic detection which mitigates many causes, and consequences of honesty violations, and helps improve strategies in handling honesty violations in mobile apps. These are summarised in Table \ref{tab:dx_findings} and explained in the subsections below. We quote interview participants by IP$<$ID$>$ and survey participants SP$<$ID$>$.

\begin{table}[]
\caption{Findings (Automatic detection mitigates causes, consequences, and improves the strategies marked in \textbf{bold}); (\#): participant count)}
\label{tab:dx_findings}
\resizebox{\textwidth}{!}{%
\small
\begin{tabular}{@{}lllll@{}}
\toprule
                                                                                          & \textbf{Business}                                                                 & \textbf{Developers}                                                                                         & \textbf{\begin{tabular}[c]{@{}l@{}}App \\ Platforms\end{tabular}} & \textbf{Users}                                                                                            \\ \midrule
\multicolumn{5}{l}{\cellcolor[HTML]{EFEFEF}\textbf{Honesty violations in mobile apps}}                                                                                                                                                                                                                                                                                                                                                                                      \\
                                                                                          & Maximise revenue (31)                                                             & Poor designing (12)                                                                                         & \textbf{Vague audits (1)}                                         & \begin{tabular}[c]{@{}l@{}}False claims \\ (competitors \\ in addition to users) \\ (7)\end{tabular}      \\
                                                                                          & Market competition (6)                                                            & Poor testing (6)                                                                                            &                                                                   &                                                                                                           \\
\multirow{-3}{*}{\textbf{Causes}}                                                         & \begin{tabular}[c]{@{}l@{}}Improper definition of \\ target audience\end{tabular} &                                                                                                             &                                                                   &                                                                                                           \\ \midrule
                                                                                          & \textbf{Bad reputation (22)}                                                      & \begin{tabular}[c]{@{}l@{}}Extra work to fix honesty \\ violations (6)\end{tabular}                         &                                                                   & Identity theft (9)                                                                                        \\
                                                                                          & \textbf{Face legal issues (8)}                                                    & \begin{tabular}[c]{@{}l@{}}Experience negative \\ emotions (7)\end{tabular}                                 &                                                                   & \textbf{\begin{tabular}[c]{@{}l@{}}Experience negative \\ emotions (21)\end{tabular}}                          \\
                                                                                          & \textbf{Lose user trust (8)}                                                      & Harm work performance (3)                                                                                   &                                                                   & \begin{tabular}[c]{@{}l@{}}Lose trust in apps/ \\ company/ developers \\ (14)\end{tabular}                \\
                                                                                          & \textbf{Lose users (7)}                                                           & Harm personal reputation (7)                                                                                &                                                                   & \begin{tabular}[c]{@{}l@{}}Lose money \\ unknowingly (19)\end{tabular}                                    \\
                                                                                          & \textbf{\begin{tabular}[c]{@{}l@{}}Lose revenue/ \\ business (18)\end{tabular}}   &                                                                                                             &                                                                   & Lose time (4)                                                                                             \\
\multirow{-6}{*}{\textbf{Consequences}}                                                   &                                                                                   &                                                                                                             &                                                                   & \textbf{\begin{tabular}[c]{@{}l@{}}Stop using/ uninstall/ \\ not install apps (13)\end{tabular}}          \\ \midrule
                                                                                          &                                                                                   & \textbf{\begin{tabular}[c]{@{}l@{}}Strengthen designing \\ practices (7)\end{tabular}}                      &                                                                   &                                                                                                           \\
                                                                                          &                                                                                   & \textbf{\begin{tabular}[c]{@{}l@{}}Strengthen development \\ practices (6)\end{tabular}}                    &                                                                   &                                                                                                           \\
                                                                                          &                                                                                   & \textbf{\begin{tabular}[c]{@{}l@{}}Strengthen testing \\ practices (20)\end{tabular}}                       &                                                                   &                                                                                                           \\
                                                                                          &                                                                                   & \begin{tabular}[c]{@{}l@{}}Be transparent with \\ customers/ users (16)\end{tabular}                        &                                                                   &                                                                                                           \\
\multirow{-5}{*}{\textbf{\begin{tabular}[c]{@{}l@{}}Avoiding \\ strategies\end{tabular}}} &                                                                                   & Have moral standards (5)                                                                                    &                                                                   &                                                                                                           \\ \midrule
                                                                                          &                                                                                   & \begin{tabular}[c]{@{}l@{}}Thoroughly investigate the \\ violation and fix (30)\end{tabular}                &                                                                   &                                                                                                           \\
                                                                                          &                                                                                   & Hotfix (17)                                                                                                 &                                                                   &                                                                                                           \\
                                                                                          &                                                                                   & \begin{tabular}[c]{@{}l@{}}Be transparent about the \\ violation with customers/ \\ users (14)\end{tabular} &                                                                   &                                                                                                           \\
\multirow{-4}{*}{\begin{tabular}[c]{@{}l@{}}\textbf{Fixing} \\ \textbf{strategies}\end{tabular}}            &                                                                                   & \begin{tabular}[c]{@{}l@{}}Have tools in place to \\ resolve honesty violations \\ (2)\end{tabular}         &                                                                   &                                                                                                           \\ \midrule \midrule
\multicolumn{5}{l}{\cellcolor[HTML]{EFEFEF}\textbf{Automatic detection of honesty violations}}                                                                                                                                                                                                                                                                                                                                                                              \\
                                                                                          & \begin{tabular}[c]{@{}l@{}}Retain/ improve \\ reputation (11)\end{tabular}        & \begin{tabular}[c]{@{}l@{}}Quick detection of honesty \\ violations (20)\end{tabular}                       &                                                                   & \begin{tabular}[c]{@{}l@{}}Transparency by \\ knowing what to \\ expect from the \\ app (15)\end{tabular} \\
                                                                                          & \begin{tabular}[c]{@{}l@{}}Reduce/ avoid \\ legal risks (4)\end{tabular}          & \begin{tabular}[c]{@{}l@{}}Improve developer \\ satisfaction (5)\end{tabular}                               &                                                                   & \begin{tabular}[c]{@{}l@{}}Find honest apps \\ in stores (4)\end{tabular}                                 \\
                                                                                          & \begin{tabular}[c]{@{}l@{}}Gain more \\ revenue (3)\end{tabular}                  & Avoid fixes (2)                                                                                             &                                                                   & \begin{tabular}[c]{@{}l@{}}Improve user \\ satisfaction (13)\end{tabular}                                 \\
                                                                                          & \begin{tabular}[c]{@{}l@{}}Retain/ gain \\ users (3)\end{tabular}                 & Reduce effort on fixing (2)                                                                                 &                                                                   &                                                                                                           \\
\multirow{-5}{*}{\textbf{Benefits}}                                                       & \begin{tabular}[c]{@{}l@{}}Improve user \\ trust (6)\end{tabular}                 &                                                                                                             &                                                                   &                                                                                                           \\ \bottomrule
\end{tabular}%
}
\end{table}

\subsubsection{Causes (RQ3.1)}
\begin{figure}
    \centering
    \includegraphics[width=\textwidth]{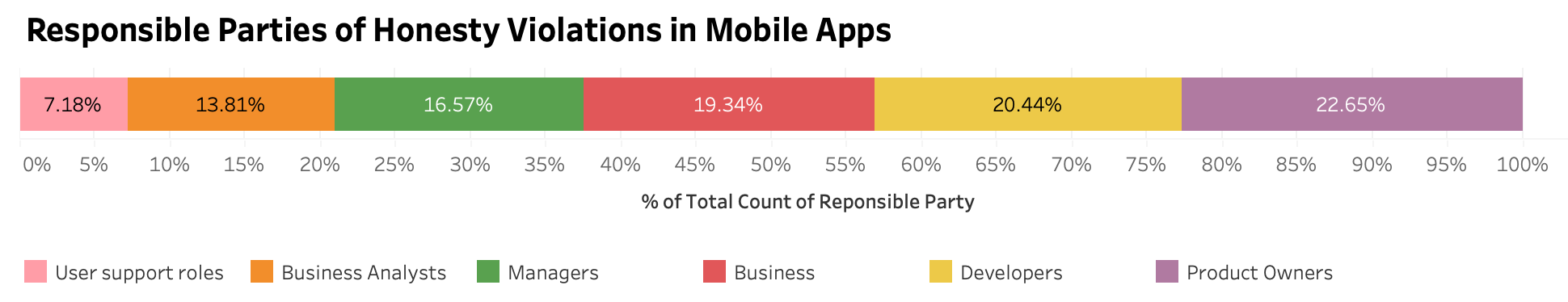}
    \caption{Responsible Parties of Honesty Violations in Mobile Apps}
    \label{fig:dx_responsible_party}
\end{figure}

QUANTITATIVE FINDINGS: 
 The majority of the participants (22.65\%) selected the choice \textit{product owners} as responsible for honesty violations in mobile apps, followed by developers (20.54\%), managers (19.34\%), business analysts (13.81\%), and user support roles (7.18\%) (Fig. \ref{fig:dx_responsible_party}). But, as the answers to the open--ended questions, and during the interviews, the experiences they shared were about businesses, developers, app platforms, users and competitors causing honesty violations (explained under qualitative findings below). However, in agile contexts, as a common practice at present, the developers are cross--functional and play multiple roles.

\noindent QUALITATIVE FINDINGS: 
Our participants identified several causes of the existence of honesty violations in mobile apps. We categorised them according to 
the driver of the violation seems to be: business drivers, developer drivers, app platform drivers, user drivers, and competitor drivers. Some honesty violations in mobile apps may of course have multiple of these drivers. \\

\underline{\noindent\textbf{\textit{Business Drivers.}}} We found intentional and unintentional reasons driven by business needs or perceived needs, for the existence of honesty violations in their mobile apps.

\textit{Intentional reasons:} Businesses intentionally violate honesty in mobile apps, as identified by our participants. The reasons for ill-intended activities mentioned include due to revenue maximisation and market competition. 

\textbf{Maximise revenue: } Our participants mentioned that when the objective of a business is to gain more profit, they may be tempted to scam/fool their app users: 
\newline \faCommentsO \textit{``I try to think these "violations" aren't meant to be there, but I did see 'tricks' to implement them, mostly: unfair fees or hidden fees, and mostly is because corporate greediness'' – SP46
}

This has become relatively easy as the smartphone user population is high and most users are not knowledgeable enough to understand the violation: 
\newline \faCommentsO \textit{It's an easy way to make money since most of the population has a smartphone and many of those users are not knowledgeable. – SP26}

As shared by SP37, tricking users into gaining more money has also led to bad practices such as doing R \& D on these tricks: 
\newline \faCommentsO \textit{``I think though that most app developers that do that kind of violations are not at all naive and do it for the money. In every app store (Apple or Google) there are tons of apps that aim to make the most money with the cheapest tricks. I feel like that a good chunk of the market is only there to milk naive consumers. I have heard stories of fellow developers/managers that pour more R \& D on how to trick people out of their money rather than put time and effort to come up with a good idea for an app and polish it'' – SP37}

\textbf{Market competition: }Due to high competition in the mobile app market, businesses may make releases that include honesty violations. These could be innovative features: 
\newline \faCommentsO \textit{``It is more and more difficult to publish applications, and the need to stand out from the competition requires to release innovative things requiring more interaction with the user (camera, localization, ...) and it is often a hinders adoption'' – SP23}

And exaggerations and misleading information: 
\newline \faCommentsO \textit{``..with all of the competition in the mobile space it is hard to stand out which makes companies feel a need to often exaggerate details to gain an edge''– SP38}

\textit{Unintentional reason:}
The reason shared by one participant is that businesses \textbf{improperly define their target audience}, which results in broad markets, and not being able to follow local laws. For an instance, if the market is too broad and the app is available for various countries and regions, the laws of particular countries or regions may not be followed specifically: 
\newline \faCommentsO \textit{``Also can happen in my opinion unintentionally: ... Market too broad, not targeting your users correctly, Not following the local laws'' – SP64}

\underline{\noindent\textbf{\textit{Developer Drivers.}}}
We found poor engineering practices were reported to cause honesty violations in mobile apps, including poor design and inadequate testing.

\textbf{Poor design: }
As shared by our participants, developers not considering honesty when designing the app functionality can result in many kinds of honesty violations in the apps. This could be because of the developers’ unawareness of the user expectations of honesty, neglecting the requirement of honesty when designing the app, and poor product analysis:
\newline \faCommentsO \textit{``This could be caused by negligence on the part of the mobile app development team when designing and developing the app hence causing it not to function properly. Also providing false information or omitting vital information.'' – SP39}

\textbf{Poor testing: }
Our participants mentioned that bugs that were not found before releasing the apps can result in honesty violations: \newline \faCommentsO 
\textit{``Probably due to malware or because of a bug during the planning or construction phase. Especially for the no service situation, most of the time it happens due to limited internet connection.'' - SP56}

\underline{\noindent\textbf{\textit{App Platform Drivers. }}}
One of our participants mentioned that \textbf{vague audits} done by app platforms allow honesty violations to exist in mobile apps: \newline \faCommentsO 
\textit{``This is because app platforms like PlayStore and others do very vague audits for these types of applications; an ethical developer will never create an app with those purposes, what he will do is provide a real service and then try to gain some benefits if possible. The blame is on these fake "developers" and digital platforms who make them public without performing a proper audit.'' – SP69}

\underline{\noindent\textbf{\textit{User Drivers.}}} 
Some of our participants mentioned that app reviews with honesty violations do not always indicate that there is an honesty violation in the app i.e., \textbf{false claims} exist, which may also be unfair to the app.
As shared by our participants, the reasons for users making false claims may include:

(a) end-user ignorance, i.e., they have not paid attention to what they signed up for: \newline \faCommentsO \textit{``I think usually they happen because people are distracted and don't pay attention to what they sign up for'' – SP55}, 

(b) faulty understandings because similar apps have honesty violations:  \newline \faCommentsO \textit{``I think people are confused and may think they are getting scammed when they aren't. There are also a lot of apps out there that are scammy so it puts the customer in a state of paranoia or fear'' – SP26},

(c) misusing their right to honesty by lowering the threshold, which seems to be unfair: 
\newline \faCommentsO \textit{``The fact that there's a general discrepancy in how much an app is worth to the user compared to a developer who puts effort into making and improving gives the users the idea that they can rightfully lower their honesty threshold'' – SP22},

(d) hate towards the purpose of the app from diverse user groups: 
\newline \faCommentsO \textit{``Once, I got a lot of bad reviews (claiming it's a scam, and similar) during a very short time period (1 day), all from Russia. It seems my app was a target of some gay hating group, since this app was a gay dating app'' – SP19}, 

and (e) claims from unintended target market: \newline \faCommentsO \textit{``In my opinion, it happens when you don't define the scope of your app well, meaning:* Intended target (age, for example)* Market. ... So you have many variables, many different laws, moral systems/opinions, cultural differences, so things that are seen as normal here are not in other places.'' – SP46}.

\underline{\noindent\textbf{Competitors.}}
One of our participants mentioned that competitors also report honesty violations through app reviews, which we think assume to be \textbf{false claims}: 
\newline \faCommentsO \textit{``People might be from other company that produces similar product'' – SP2}

\subsubsection{Consequences (RQ3.2)}
We asked the participants about the impact of honesty violations in apps on owners (businesses), developers, and users. From what the participants shared, we found the consequences of honesty violations in mobile apps.

\underline{\noindent \textit{\textbf{Business.}}} 
Bad reputation, facing legal issues, loss of user trust, loss of users, and loss of revenue/business are the consequences of honesty violations in apps on businesses as we found.

\textbf{Bad reputation:} The majority of the participants mentioned that if honesty violations exist in the apps, that will create a bad reputation for the business. For example, the businesses may look like they are frauds, incompetent, and dishonest: 
\newline \faCommentsO \textit{``If the violation is found, it's a huge setback to business reputation'' -- SP25}



One participant mentioned that if the violation is intentional, then the business deserves a bad reputation: 
\newline \faCommentsO \textit{``It could worsen their reputation, but again if they happen to do fraudulent acts they kind of deserve having their reputation down.'' -- SP12}

Bad reputation will also lead to the rest of the consequences discussed below.

\textbf{Face legal issues:} As shared by some participants, users may take legal actions in some cases. As mentioned by SP64, businesses may get sued for not following the law of the country. Therefore, businesses need to be mindful and follow the laws according to the country/region before releasing the product: 
\newline \faCommentsO \textit{``Getting removed from the store, having to make urgent changed, getting sued if the laws of the country are not being followed.'' -- SP64}

The impact if the found violations are not fixed could be heavy, as shared by SP68: 
\newline \faCommentsO  \textit{``They could impact them heavily if they don't act on solving possible problems that caused those violations in the first place especially where there is relative legislations.'' -- SP68
}

\textbf{Lose user trust:} If honesty violations are found in apps, then the users might not trust the app/business anymore. This could even lead to not trusting the future apps of the company as well: 
\newline \faCommentsO \textit{``It could impact the owners of the mobile app because their customers would end up not trusting other future apps from the owner'' -- SP40
}

Even if the violations are fixed, regaining the trust back from the users is difficult:
\newline \faCommentsO \textit{``no matter how much developers try to fix their issues when users have lost trust it's hard to regain it back they move to other apps which is bad for developers because creating an app takes time and money'' -- SP3}

\textbf{Lose users: }As shared by the participants, the businesses may lose users if honesty violations are found in the apps. The users could be existing users or new users. The users may not download or subscribe/re-subscribe to the services in the app or switch to other apps. Sometimes, the users might discuss the violations in public, which will resist new users from installing the app: 
\newline \faCommentsO \textit{``It could impact downloads and/or subscriptions or re-subscriptions.'' -- SP58}



\textbf{Lose revenue/business: }All the above-mentioned consequences may lead to revenue loss and in extreme cases where the app gets removed from the store, businesses may lose their businesses too as shared by our participants:
\newline \faCommentsO \textit{``The developers or owners of the mobile apps can be impacted by negative reviews, which would lead to decreased sales'' -- SP62}



\underline{\noindent \textit{\textbf{Developers. }}}
Extra work to fix honesty violations, experience negative emotions, harm work performance and harm personal reputation are the consequences of honesty violations on developers as we found.

\textbf{Extra work to fix honesty violations: }Irrespective of the claim being true or false, the developers will have to work extra. Additional tasks may need to be done to look into the reported violations. If the app reviews are false, then as shared by our participants, the developers will lose time unnecessarily by looking for a violation which does not exist in the app: 
\newline \faCommentsO \textit{``The developers will be confronted with additional tasks and will try to search for eventual bugs mentioned by the users. In case of dishonest reviews, the app developers may lose time looking for bugs that don't even exist.'' -- SP66}

\textbf{Experience negative emotions:} As mentioned by our participants, the developers experience negative emotions such as stress and anger when they receive honesty violation reported reviews. They also feel guilt about having the app developed with honesty violations:
\newline \faCommentsO \textit{``Developers and owners can experience a lot of stress.'' -- SP1}

\textbf{Harm work performance:} Three participants mentioned that finding honesty violations damage the work performance of the developers. They could get demotivated to work, which harms their performance: 
\newline \faCommentsO \textit{``maybe they are ashamed and in the long run it kills motivation'' -- SP38}

\textbf{Harm personal reputation: }Our participants also stated that finding honesty violations in the apps developers develop impacts their personal reputation negatively, which is problematic for the career of the developers: \newline \faCommentsO \textit{``A bad thing that could happen to developers is the result of negative reviews they receive from customers, thus resulting in a declining career.'' -- SP70}

Also, SP67 stated that it is common for developers to get the blame as they are at the bottom of the tree of the team structure: 
\newline \faCommentsO \textit{``It could affect the developers because they are usually the ones to get the blame for these violations. The blame just gets pushed down the tree until it hits the developers that just have to take it.'' -- SP67}

\underline{\noindent \textit{\textbf{Users.}}} 
Identity theft, experiencing negative emotions, loss of trust in apps/company/developers, loss of money unknowingly, loss of time, stop using/uninstalling/not installing apps are the consequences of having honesty violations in apps on users.

\textbf{Identity theft:} One of the concerns our participants mentioned (as an impact on end users due to the existence of honesty violations in mobile apps) is identity theft. The apps with honesty violations could compromise personal and sensitive data. For example, their home address and credit card details. These could be used by third parties in various activities. This could even lead to frustration in users, affecting their mental health as well: 
\newline \faCommentsO \textit{``It could make the end users prone to fraudulent activities on their finances by making use of their personal information to extract money or some resources from them without their consent. It could also make them prone to some physical attacks such as the end users' home address being exposed, thereby making their locations traceable.'' -- SP40}

\textbf{Experience negative emotions:} If honesty violations are found in mobile apps, users may experience negative emotions as shared by our participants. These include frustration, unhappiness, stress, and anger: 
\newline \faCommentsO \textit{``End users become very frustrate due to these honesty violations.'' -- SP41}

\textbf{Lose trust in apps/company/developers: }As stated by our participants, when users find honesty violations in the apps they use, no matter how legit the brand of the company, users may end up not trusting and distancing themselves from the app and company: 
\newline \faCommentsO \textit{``They create a feeling of lack of trust with the company responsible for the violations'' -- SP65}

This is more related to cancellations and refund policies:
\newline \faCommentsO \textit{``if we look at cancellations and refund policies, I think everyone as if you're a user of a mobile app, you would want to be able to trust the app that you're using. And especially if you're paying money for it, for example, a subscription service or even just a one-time payment. If it's a subscription model then you would want to be able to save the cancel your subscriptions without having any issues. If the app that you're using doesn't actually provide that sort of honesty, capabilities like then you and I personally wouldn't be comfortable using an app that was an honest to their users'' -- IP3}

\textbf{Lose money unknowingly: }Users losing money without them knowing is one of the key impacts of honesty violations claimed by our participants. This could also lead to a problematic living in physical life as money plays a huge role: 
\newline \faCommentsO \textit{``Honesty violations can impact users in various ways. In my opinion, the worst effect it can have is when a person loses money that they weren´t expecting to lose, it can very realistically impact their ability to buy essentials - pay rent - pay bills - etc, if not refunded.'' -- SP38}

\textbf{Lose time: }Similar to losing money, losing time is due to the existence of honesty violations can cause negative effects on the lives of people by wasting the valuable time they could use to do other important work: 
\newline \faCommentsO \textit{``People will be made addicts to apps and forced to spend huge amounts of time, which they could use for more productive work.'' -- SP34}

\textbf{Stop using/uninstall/not install apps: }When users find honesty violations in apps, they are likely to either stop using them or uninstall them. If any potential new users get to know the existence of the violations in the apps, they might not even install them: 
\newline \faCommentsO \textit{``It could scare them away from using the app or ever downloading it. This would be very negative for everyone involved if the app does do what it is supposed to do.'' -- SP67}

\subsubsection{Strategies (RQ3.3)}
We explored strategies that developers use to avoid honesty violations, and strategies the developers use to fix honesty violations when they receive feedback containing honesty violations for the app.

\underline{\noindent \textit{\textbf{Mitigating strategies.}}} The mitigation strategies include having better engineering practices, being transparent with the customers/users, and having moral standards.

\textbf{Having better engineering practices:} We found that having better engineering practices in place helps software teams to avoid honesty violations to occur in the apps they develop. This includes strengthening design practices,  development practices, and the testing process.

(a) Strengthening design practices: Viewing the app from the user perspective helps in avoiding honesty violation occurrences in the apps being developed, as reported by participants: 
\newline \faCommentsO \textit{``Keep clear communication with your manager, Keep on top of the timeline that is given, explain how users will see the app to your manager.'' -- SP67}

Having honesty as a first-class app requirement, and complying with regulations/policies (e.g.: GDPR) also helps to mitigate occurrences of honesty violations in apps according to our participants:
\newline \faCommentsO \textit{``...Other than that, we gotta respect the GDPR.'' -- SP29}

(b) Strengthening development practices: Some participants mention that they have codes in place to flag honesty violations:
\newline \faCommentsO \textit{``We also have codes in place which will flag an app in violation'' -- SP14}

They also believe that having security measures in place avoids honesty violations from happening. For example, as SP18 mentioned, the violations could occur by attacks on the app, so they use security measures accordingly. Measures such as pen tests could avoid risks, and encryption could avoid data leakages and thefts: 
\newline \faCommentsO \textit{``Of course as always we use source code encryption. Attackers generally repack renowned apps into rogue apps using reverse-engineering techniques. Then they upload those apps into third-party app stores with the intent to attract unsuspecting users. It has been a consistently good practice to test our application against randomly generated security scenarios before every deployment. Especially, pen testing can avoid security risks and vulnerabilities against our mobile apps. Detecting loopholes in the system is an absolute necessity. Since these loopholes could grow to become potential threats that give access to mobile data and features. The sensitive information that is transmitted from the client to the server needs to be protected against privacy leaks and data theft. When it comes to accessing confidential data, the mobile apps (yes, we have more than one app in our company)  are designed in a way that the unstructured data is stored in the local file system and/or database within the device storage. We use encryption methods like AES with 512-bit encryption, 256-bit encryption and SHA-256 for hashing. We have security measures in place to safeguard against malicious attacks at backend servers.'' -- SP18}

(c) Strengthening the testing process: A strong pre-release testing process is necessary. As shared by our participants, getting the QC teams to do stringent reviews, and having multiple testing rounds helps to avoid having honesty violations in apps after release: 
\newline \faCommentsO \textit{``
The app goes through multiple rounds of testing to make sure such violations do not happen. -- SP25}

Through our analysis, we also found that testing for honesty violations during UAT and testing the app with diverse users helps to mitigate honesty violations. Approaches such as sending the beta version of the app to users, and asking a variety of users to test the app help in looking at the app from different angles, which eventually helps in identifying honesty violations early. As shared by one participant, having users involved in testing before releasing could help in identifying 50\% of the problems with regard to honesty violations: 
\newline \faCommentsO \textit{``Before I release the application, to say general public, you, you might want to test it out on a certain set of users and look into the things that they feedback on and try and figure out where all whether the feedback that they give violates any of these honesty statements, and from there onwards you might be able to improve on specific ones that are being violated. But if, yeah, that would be the initial thing. Then once I think from there, you can sort of read out maybe like 50\% of the problems regarding honesty violations. Then maybe once you've actually released your application to the general public, then from there on, you would have a much larger audience size and also a lot more feedback that you can work on.'' -- IP3}

\textbf{Being transparent with customers/users: }
Being more transparent with customers/users helps identify and address honesty violations ``reported'' in the apps as reported by our participants. For example, they commented that having additional confirmation steps, and being transparent in the app description helps well informing the users about the app: 
\newline \faCommentsO \textit{``Present the end user with correct information and make them do additional confirmation steps.'' -- SP57}

\textbf{Having moral standards: }
Being a person who values things such as honesty, fairness, and credibility helps to avoid honesty violations in mobile apps, as shared by our participants: 
\newline \faCommentsO \textit{``We are simply not designing and developing apps in such a way. Honesty and integrity is key'' -- SP8}

\underline{\textit{\textbf{Fixing strategies.} }}
The fixing strategies shared by our participants include thoroughly investigating honesty violations flagged by users and fixing, hot-fixing, being transparent about the violation with customers/users, and having tools in place to help resolve honesty violations.

\textbf{Thoroughly investigate the violation and fix: }
The majority of our participants mentioned that they thoroughly investigate any reported honesty violation by users and then move forward with fixing it. For that, they said that they have team discussions, consult supervisors, and even get more developers involved in fixing: 
\newline \faCommentsO \textit{``I usually consult my supervisors on how to best go about handling such situations if and when they occur.'' -- SP6}

One participant mentioned, they talk with the legal department as well before fixing, and then change the code: 
\newline \faCommentsO \textit{``Talk with the legal department and change the code respectively.'' -- SP68}

Thorough investigation includes checking the reliability of the claims, checking the alignment of the claim with business policies -- which sometimes hinders fixing the violation, internally validating the policy applied, removing the app temporarily from the app store, testing the app thoroughly, and adding to the sprint, prioritise, and then fix: 
\newline \faCommentsO \textit{``I have to consult the architect to discuss the solution. Sometimes it's not possible to fix it due to business policies'' -- SP25}

\textbf{Hotfixing: }
Some participants mentioned that they do `hotfixes', i.e., they fix the app honesty violation immediately: 
\newline \faCommentsO \textit{``If I had any, I would immediately make the violation in question disappear by fixing the violation.'' -- SP35}

Some stated that hotfixes depend on the size of the company. For example, if the app is owned by a solo developer/small company, then they will fix it quickly rather than large established companies which may have long processes: 
\newline \faCommentsO \textit{``If you are the solo developer of your own application,  then you can address it pretty quickly. So you got your feedback, then you probably can use that feedback to generate a list of things that you can improve on. Then there will be a pretty quick turnaround based on user feedback, but usually on larger company sizes. There's a lot of feedback and you probably want to address the feedbacks that are more serious. And so that would take some time and then that's usually a really long process before it gets like from gathering the requirements to putting it into user stories and having developers work on it and then push to production might take potentially months. So it's not a quick process. If it's a big company, but obviously it depends on how the company works.'' -- IP3}

\textbf{Be transparent about the violation with customers/users: }
We found that our participants consider being transparent about the issue and changes they make in the app is important. They mentioned that they inform their users through app descriptions about temporal deactivations due to technical issues, and also they do the same after fixing by sharing what was fixed. However, we could not find if they mention the type of honesty violation in their temporary app description updates. Some also mentioned that they share a report with full transparency with their users/customers: 
\newline \faCommentsO \textit{``Summarily, I put out a notice to users that there'll be a temporal downtime in order to address technical issues that could affect their use of the app after I do this, I temporarily deactivate the app and identify the vulnerability, write a code that can patch it up, then I activate it and announcer to the users that the app is running again.'' -- SP40}

\textbf{Have tools in place to resolve honesty violations.:}Two of our participants mentioned that they would have measures/tools implemented to resolve the honesty violations: 
\newline \faCommentsO \textit{``My team investigates them and implements tools to resolve them.'' -- SP39}

\subsubsection{Usefulness and Benefits of Automatic Detection of Honesty Violations (RQ 3.4)}

\begin{figure}
    \centering
    \includegraphics[width=\textwidth]{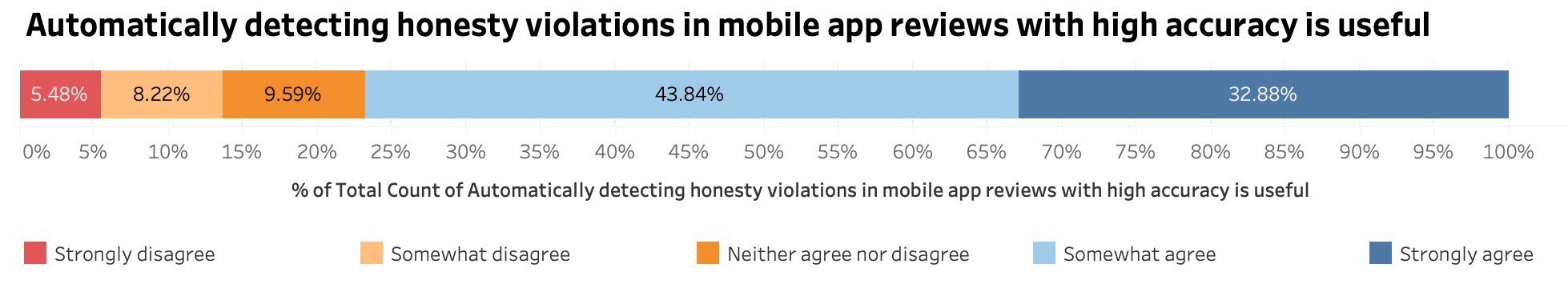}
    \caption{Participants' Opinion about Automatic Detection of Honesty Violation in Mobile App Reviews}
    \label{fig:dx_usefulness}
\end{figure}

\paragraph{USEFULNESS.}
Our participants shared their opinion about the automatic detection of honesty violations in mobile app reviews (Fig. \ref{fig:dx_usefulness}). Note that, participants did not use our tool to answer this question, but shared their general opinion about the automatic detection of honesty violations in mobile app reviews. The majority of the participants somewhat agreed (43.84\%) that automatically detecting honesty violations in mobile app reviews with high accuracy is useful, then 32.88\% strongly agreed with it, and the rest neither agreed nor disagreed/somewhat disagreed/strongly disagreed.

\begin{figure}
    \centering
    \includegraphics[width=\textwidth]{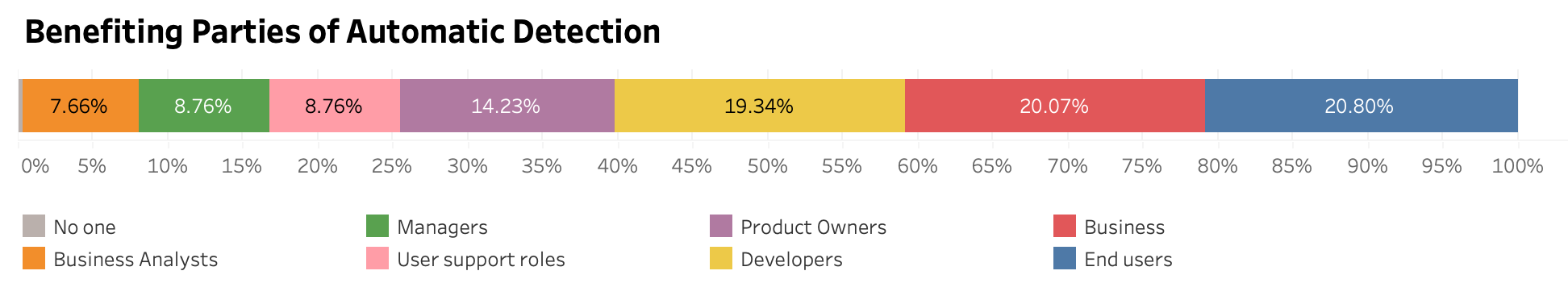}
    \caption{Benefiting Parties}
    \label{fig:dx_benefiting_parties}
\end{figure}

\noindent \paragraph{BENEFITING PARTIES.}
According to our participants, end users (20.80\%) will be majorly benefited from automatic detection, followed by businesses (20.07\%), developers (19.34\%), product owners (14.23\%), managers and user support roles (8.76\% each), and business analysts (7.66\%). 0.36\% of the responses were for \textit{no one} too. This is summarised in Fig. \ref{fig:dx_benefiting_parties}. However, from the qualitative data, we found businesses, developers, app platforms, and end users have the potential to get the most out of automatic detection.

\noindent \paragraph{BENEFITS.}
We asked our participants how automatic detection of honesty violations could be beneficial. They shared a variety of benefits which we categorised as benefits for businesses, developers, and users. In this section, we use the term ``automatic detection'' as a short term for ``automatic detection of honesty violations''.

\underline{\textit{\textbf{Business.}}} Retain/improve reputation, reduce/avoid legal risks, gain more revenue, retain/gain users, and improve user trust are the benefits for businesses from automatic detection.

\textbf{Retain/improve reputation: }As automatic violation fastens the honesty violation fixing process, businesses are able to fix them early before that can negatively impact their reputation, and also users may leave positive user reviews which will improve the businesses' reputation: 
\newline \faCommentsO \textit{``They will be aware of the violations in time and fix them before it can ruin their reputation'' -- SP9}

\textbf{Reduce/avoid legal risks: }Since automatic detection helps improve the fixing process in the software team, the team will be aware of the honesty violations, and they will take action to fix the issues in a timely manner, which will avoid any legal issue occurrences, and if any risks were there, those will get reduced too as shared by our participants: 
\newline \faCommentsO \textit{``Just avoiding further issues with the law, and economic backlash.'' -- SP64}

\textbf{Gain more revenue:} Happy customers bring better revenue as shared by our participants. Better user experience, and users enjoying the app as automatic detection helps fix issues faster, leading to an increase in revenue. Not only that but leaving better user reviews will also help in increasing sales: 
\newline \faCommentsO \textit{``The consumers will benefit from having a better experience through the app, and the developers and owners of the app will, in turn, benefit from satisfied customers leaving better reviews, resulting in increased sales and a good reputation.'' -- SP62}

\textbf{Retain/gain users: }Since automatic detection improves the app and how users see it, businesses will not lose users and also will gain more new users as shared by our participants. This is interconnected to gaining more revenue -- as not losing any users and gaining new users mean better revenue, and a good reputation -- as users will be retained with the business because of the good reputation of the business: 
\newline \faCommentsO \textit{``I think they will benefit in a sense that once the reviews are there it can be fixed with immediate effect, this resulting in good customer care service as well as prompting user to give good/positive reviews which could lead to new clients in the end, which benefits the app/business in the end.'' -- SP17}

\textbf{Improve user trust: }As an app will be improved by having automatic honesty violation detection, users will trust the app as shared by our participants: 
\newline \faCommentsO \textit{``The app will improve and other users are going to trust in the app and download it'' -- SP36}

\underline{\textit{\textbf{Developers.}}} Through our analysis, we found the potential benefits the developers may have due to automatic detection. The benefits for developers are quick detection of honesty violations, improve developer satisfaction, avoiding fixes, and reduced effort on fixing.

\textbf{Quick detection of honesty violations:} Having the detection of honesty violations automated could benefit the developers by fastening the detection process. They will spend less time doing investigations, and will focus on fixing more: 
\newline \faCommentsO \textit{``For the owners and developers of the app, it will be easier to sort them out and do what they have to do.'' -- SP12}

\textbf{Improve developer satisfaction: }As their work gets easier and less hectic, the developer satisfaction will be benefited from automatic detection as shared by our participants: 
\newline \faCommentsO \textit{``And also having a good state of mind.'' -- SP46}

\textbf{Avoid fixes: }If honesty violations are found early and fixed early, that will avoid future fixes as mentioned by our participants. Because of this, the developers may use their time to improve the app rather than spending time on fixing: 
\newline \faCommentsO \textit{``Developers and product owners can employ their time in improving the app instead of fixing it.'' -- SP42}

\textbf{Reduce effort on fixing: }The developers' time spent on the entire fixing process including reading the app reviews, analysing them, and then moving forward with fixing will be reduced by having automatic detection involved in the process, as stated by our participants:
\newline \faCommentsO \textit{``Automation will create less work for them. Also, it will take automatic action rather than slow and expensive human interference.'' -- SP53}

\underline{\textit{\textbf{App platforms.}}} One of the participants mentioned that app platforms could benefit if they use automatic detection.

\textbf{Improve audits:} One of the causes found for the existence of honesty violations in mobile apps is vague audits by app platforms. This could possibly be improved if the app platforms use automatic detection to examine apps with several app reviews reported with honesty violations: 
\newline \faCommentsO \textit{``I think that the one group that would benefit the most is end users. App stores might be finally cleaned up from scummy apps and trashy developers.'' -- SP37}

\underline{\textit{\textbf{Users.}}} Not only businesses, developers, and app platforms can be benefited from automatic detection, but also users as mentioned by our participants. The benefits to users as found are, transparency by knowing what to expect from the app, finding honest apps in stores, and improve user satisfaction.

\textbf{Transparency by knowing what to expect from the app: }As stated by our participants, if automatic detection is available for users, they may also use it to better understand the app. A high level of transparency will be available through it and because of that users will know what to expect from the app -- whether to expect honesty violations in apps or not. This will help users in deciding whether to download to an app or not: 
\newline \faCommentsO \textit{``They  will have a framework which guides and informs them regarding the apps they are using/working .'' -- SP14}

\textbf{Find honest apps in stores: }If the businesses, developers, and app platforms take the maximum use of automatic detection, the users will obviously find honest apps in stores as shared by our participants: 
\newline \faCommentsO \textit{``They will limit the ability of some companies to mock customers and try to curry favour with them. In addition, they will guide end users not to use apps that cheat them.'' -- SP67}

\textbf{Improve user satisfaction:} The user satisfaction will be improved if the software teams use automatic detection as apps will be improved which will give less frustration to the users and better user experience, which result in better user satisfaction: 
\newline \faCommentsO \textit{``Customers won´t feel as if they are being lied to and so they will have a much happier experience with the product, and therefore customer support will have fewer unhappy customers to deal with.'' -- SP38}

\begin{tcolorbox}[boxrule=1pt, colback=gray!10!white, boxsep=0pt, top=3pt, bottom=3pt, left=3pt, right=3pt, before skip=3pt,after skip=3pt]
\small
\textbf{\textit{RQ3 Answer:}} The developer study revealed seven causes, seven responsible parties, sixteen consequences of honesty violations, nine strategies developers use to avoid/fix the honesty violations, and thirteen benefits and nine benefiting parties of automatic detection of honesty violations in mobile apps. 
\end{tcolorbox}

\section{Discussion and Recommendations}
\label{sec:discussion}


\subsection{Technology (Mobile Apps) as Values Artefacts}
Software artefacts such as mobile apps, like other technological artefacts, express human values \cite{Whittle:2019Is}. Although less formally articulated, human values may be reflected throughout the different phases of the software development life cycle \cite{Nurwidyantoro:2021towards}. Values are represented in the conception and abstraction of ideas, in the way software features are arranged, described and even implemented and these embodied values are typically those of their creators, e.g., software developers and other stakeholders \cite{Lennox:2020}.

Some studies have argued that technological artefacts are value-agnostic tools that can be used for good or bad (i.e., theory of the social determination of technology) \cite{Bernward:1999}, while others contend that technological artefacts are not value-agnostic, i.e., they hold value qualities and promote certain values over others \cite{Winner:1980}, e.g., the bitcoin blockchain technology \cite{nakamoto2012bitcoin} is an embodiment of the value category of self-direction. Irrespective of the sociotechnological stance on values in technological (software) artefacts, there is an agreement on the role of software artefacts in changing habits in people and influences society in general, despite the intentions of the software companies behind these artefacts \cite{Obie:2020,Agre:1997}. Sullins writes, "Since the very design capabilities of information technology influence the lives of their users, the moral commitments of the designers of these technologies may dictate the course society will take and our commitments to certain moral values will then be determined by technologists" \cite{Sullins:2018}.

Furthermore, while we do not conflate values with ethics (values are the guiding principles of what people consider in life \cite{Rokeach:1973} while ethics are the moral expectations that society agrees upon to decide which values are acceptable or not \cite{Whittle:2019Is}), the value of honesty is an ethically desired value in most societies. Thus we argue for a conscious effort in developing honest software artefacts, including mobile apps, and the promotion of honesty in software development practices. Our intention in this paper is not to serve as moral arbiters of values in mobile apps (or other software artefacts) but rather to promote a healthy discussion of these issues in the software research and development community, and point the field towards a critical technical practice of mobile SE, i.e., the reflective work of sociocultural criticisms, highlighting the hidden assumptions in technical processes, and the interaction between the social, cultural and technical aspects of (mobile) SE.

\subsection{The Role of App Distribution Platforms}
App distribution platforms such as the Apple store and the Google Play store have an important role to play in supporting the values and minimising their violations in apps published on their platforms. They can serve as enforcers of ethical systems supporting values such as honesty, akin to the manner in which they protect end-users' devices from malicious apps~\cite{li2015potential,li2017automatically}. For instance, they can ensure that app developers are transparent in their billing process and enforce a mandatory multi-step (at least two steps) confirmation not only for subscriptions but also for in-app purchases.

Another issue on the violation of the value of honesty is related to the non-transparency in the subscription process in apps. For example, while some apps provide a reminder to the user before the end of a trial period so the user can decide to cancel their subscription or progress to a premium service, some other apps provide no reminder whatsoever. A reminder-to-cancel (or upgrade) feature for apps can be necessitated by the distribution platforms to protect the end-user from unintentional subscriptions.

In addition, for games or game-like apps involving the use of money for play, end-users perceptions of unfairness in these systems can be assuaged by a practice of auditing the systems to ensure statistical outcomes that are not only probable but fair to both the end-user and app developers alike, similarly to the way casino systems are routinely audited for fairness and transparency. The results of the audits can then be shown as part of the app information on the app stores.

\subsection{Transparent Policies and Agreements}
In cases of disputes between end-users and app vendors, where an end-user perceives that they have been unfairly treated, it is typical for the app vendors to refer the end-user to the end-user licence agreement (EULA) signed by the end-user during installation \cite{KingChelsea2017Fptw}. A EULA is a legally binding contract between the end-user and the app vendor \cite{CheeFlorenceM2012RREE}.

Some app vendors place their data handling and billing processes in the fine print of EULAs that are typically difficult to understand by the average user because they are written in legal terms \cite{KingChelsea2017Fptw}. Some studies have also shown that most end-users  who clicked “I agree” do not understand the terms to which they agreed and often expressed genuine concern when the terms are expressed to them \cite{CheeFlorenceM2012RREE}. Thus it is important to develop transparent legal policies and easy-to-comprehend EULAs to inform and empower the end-user and help them understand the terms and implications of these kinds of legal contracts. Transparency and comprehensibility would alleviate wariness and misgivings in this area.  Also, we reiterate the position of O’Neill \cite{Oneill:2002}, that while transparency may undo secrecy, “it may not limit the deception and deliberate misinformation that undermine relations of trust. If we want to restore trust we need to reduce deception and lies, rather than secrecy” \cite{Oneill:2002}. This area is particularly ripe for interdisciplinary research between the computing sciences, humanities, and law. 

\begin{figure}
    \centering
    \includegraphics[width=\textwidth]{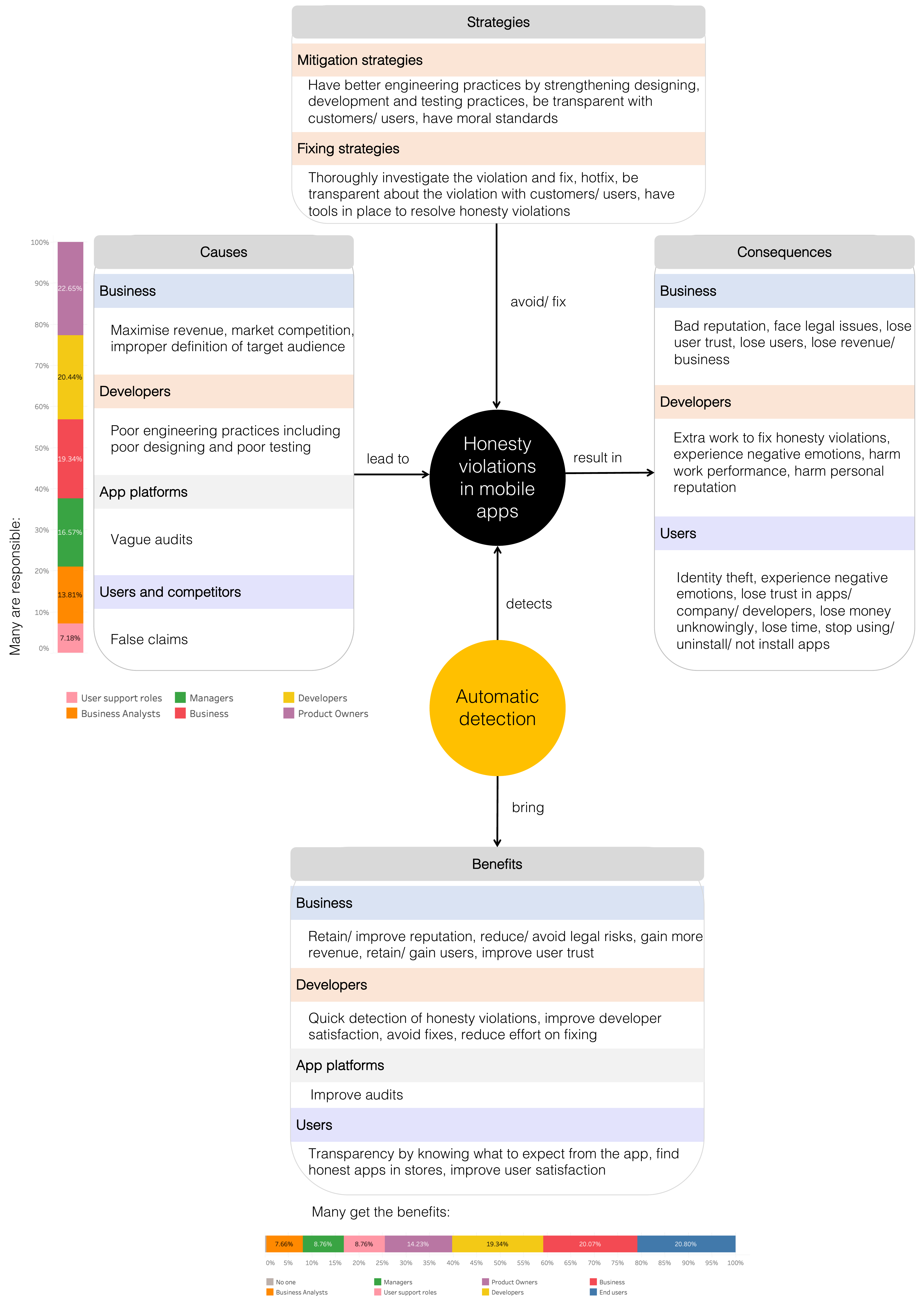}
    \caption{Causes and Consequences of Honesty Violations in Mobile Apps, Common Strategies Developers Use to Avoid/ Fix Them, and The  Automatic Detection Benefits}
    \label{fig:dx_framework}
\end{figure}

\subsection{An Actionable Framework for Developers}
Fig. \ref{fig:dx_framework} shows an actionable framework we have developed from the findings of our investigation of the developer perspective of honesty violations in mobile apps and the potential benefits of automatic detection. Almost every stakeholder involved in mobile app development is responsible for honesty violations in mobile apps, and honesty violations affect individual developers, users, and businesses at various levels. Developers use various avoiding strategies to mitigate the occurrence of honesty violations in the apps they develop, and if found any after releasing them to the public, they use a variety of fixing strategies. Automatic detection has the potential to avoid certain causes, consequences, and support developer strategies (Table \ref{tab:dx_findings}). While a minority believe no one will get benefits from automatic detection, it will be beneficial to multiple stakeholders involved, including end users, development team members, businesses, and app platforms. 

The framework could be used as a guide to get a deeper understanding of honesty violations in mobile apps. We believe that this understanding will help developers immensely to be responsible for developing mobile apps. The developers may also consider using the avoiding and fixing strategies given in this figure, as these were shared by many developers worldwide. Further, mobile app developers could consider using automatic detection to improve their internal processes, the app, and the end-user experience, which eventually helps in improving the business.

Researchers may consider exploring how various roles in teams (as in Fig. \ref{fig:dx_framework}) are responsible for honesty violations in mobile apps. They also may consider investigating the impact on those roles as we only focused on businesses, developers, and users. Similarly, as team roles have the potential to get many benefits from automatic detection, a fruitful area to investigate will be the benefits of automatic detection. Ultimately, all these findings could improve Fig. \ref{fig:dx_framework} to give the public and development teams a broad, yet in-depth knowledge of honesty violations in mobile apps, how important it is to diminish them from mobile apps, and how technology can aid this.

\subsection{Human Values in SE Research}

Research in the broader area of human values in SE is still in its early stages \cite{Perera:2020}. While the investigation of well-known values such as privacy and security has been considerably developed, other values such as honesty, curiosity, and independence have received little attention, possibly due to the subjective and abstract nature of these concepts. This and other recent related works are based on an adaptation of the Schwartz theory of basic human values \cite{Schwartz:1992}. However, the nascent field of human values in SE may benefit from new conceptual theories of human values that are more situated closely within SE.

Furthermore, there is a need for the development of tools and techniques, not only in detecting the violation of human values in software artefacts but also providing automatic recommendations for possible fixes. Directions for future work may include the following: the development of approaches for generating end-user comprehensible EULA templates supporting values, approaches for evaluating and auditing fairness in games and game-like systems to support statistically probable results, and modules for static and dynamic analysis tools to detect specific  values defects. Another area worth investigating is the development of tools for supporting the inclusion of values throughout the software development lifecycle and the resulting software artefacts, including mobile apps.

\section{Threats to Validity} \label{sec:ThreatsValidity}

\textbf{Internal Validity: } The qualitative process of building the \textbf{honesty}\textbf{\_}\textbf{discussion} dataset in Section \ref{sec:dataset}, categorising the different types of honesty violations in Section \ref{sec:approachRQ2}, categorising causes, consequences, strategies of honesty violations and benefits of automatic detection in Section \ref{sec:RQ3} might be biased and error-prone. Hence, it might have introduced some threats to the internal validity of the study. We used three techniques to mitigate such threats. First, the qualitative analysis was conducted iteratively over an ample timeframe to avoid fatigue. Second, each review was analyzed by one analyst and validated by at least one other analyst, followed by several meetings between the analysts to resolve any disagreements and conflicts; \textcolor{black}{and interview and survey findings were analysed by one analyst and shared among other analysts during weekly meetings.} Third, the analysts have extensive research experience in the area of human values.

\textbf{Construct Validity: } The analysts might have had different interpretations on the definition of the value of honesty. 
Our strategy to minimise this threat was making sure the analysts carefully examined seminal papers \cite{Schwartz:1992,Schwartz:2012} on the Schwartz theory, formal definition of honesty from dictionaries, and existing software engineering research on human values, including honesty \cite{Obie:2020,shams2020society}. In this study, among many options, we used five machine learning algorithms to detect honesty violation reviews and four metrics to evaluate the algorithms. Peters et al. \cite{peters2017text} claim that it is impracticable to use all algorithms in one study. Hence, we accept that applying other machine learning algorithms to our dataset may lead to different performances. The metrics precision, recall, accuracy, and F1-score used in this study are widely applied and suggested to evaluate machine learning models in software engineering.


\textbf{External Validity: } Our initial sample of app reviews was 236,660 reviews collected from \cite{Eler:2019} and \cite{Obie:2020}, which was further reduced to 4,885 honesty-related reviews after applying the keywords filter. Our keyword filter may have introduced false negatives and potentially excluded honesty violations in the larger dataset.
Hence, we cannot claim that our results are generalisable to all app reviews in the Apple App Store and Google Play Store and other platforms (e.g., online marketplaces). Similarly, the sample size of the developer study does not guarantee generalizability as the number of participants is limited (especially interview participants), and the majority represented Europe.
\label{threats}

\section{Conclusion} \label{sec:Conclusion}
Mobile software applications (apps) are very widely used and applied and hence need to reflect critical human value considerations such as curiosity, freedom, tradition, and honesty. The support  for -- or violation of -- these critical human values in mobile apps has been shown to be captured in app reviews.
In this work, we focused on the value of honesty. We presented an approach for automatically finding app reviews that reveal the violation of the human value of honesty from an end-user perspective. In developing our automated approach, we evaluated seven different algorithms using a manually annotated and validated dataset of app reviews. Our evaluation shows that the Deep Neural Network (DNN) algorithm provides higher accuracy than the other algorithms in detecting the violation of the value of honesty in app reviews, and also surpasses a baseline random classifier with an F1 score of 0.921.
We also characterised the different kinds of honesty violations reflected in app reviews. Our manual qualitative analysis of the reviews containing honesty violations resulted in ten categories: unfair cancellation and refund policies, false advertisements, delusive subscriptions, cheating systems, inaccurate information, unfair fees, no service, deletion of reviews, impersonation, and fraudulent-looking apps.
\textcolor{black}{We used surveys and interviews to investigate the developer perspective of honesty violations in mobile apps and how automatic detection might be beneficial. This resulted in the identification of a wide range of causes, consequences, strategies for avoiding and fixing honesty violations, and potential benefits of automatic detection. }
The results of our study highlight the importance of considering software artefacts, such as mobile apps, as an embodiment of human values with consequences on end-users and society as a whole. We emphasise the role of app distribution platforms in supporting human values, such as honesty, on their platforms, recommendations for developers, and discuss the need for the software engineering research community to investigate methods and tools to better minimise the violation of human values in software artefacts.

\begin{acknowledgements}
This work is supported by ARC Discovery Grant DP200100020. Grundy is supported by ARC Laureate Fellowship FL190100035.
\end{acknowledgements}

\bibliographystyle{spmpsci}      
\bibliography{references.bib}   

%
%

\end{document}